\documentclass[twocolumn,prb,superscriptaddress,floatfix]{revtex4-2}

\usepackage{graphicx,dcolumn,bm,amsmath,xcolor,threeparttable}
\usepackage[normalem]{ulem}

\def\bk{\mathbf k}
\def\bq{\mathbf q}
\def\bG{\mathbf G}

\def\br{\mathbf r}

\def\bQ{\mathbf Q}

\def\cms{cm$^{2}$/Vs}
\def\pcc{cm$^{-3}$}
\def\aibte{$ai$\text{BTE}}

\newcommand{\Nuc}{N_{\text{uc}}}
\newcommand{\Vuc}{\Omega_{\text{uc}}}

\begin{document}

\title{\textit{Ab initio} calculation of carrier mobility in semiconductors \\ including ionized-impurity scattering}

\author{Joshua Leveillee}
\affiliation{Oden Institute for Computational Engineering and Sciences, The University of Texas at Austin, Austin, Texas 78712, USA}
\affiliation{Department of Physics, The University of Texas at Austin, Austin, Texas 78712, USA}
\email{fgiustino@oden.utexas.edu}
\author{Xiao Zhang}%
\affiliation{Department of Materials Science and Engineering, University of Michigan, Ann Arbor, Michigan, 48109, USA}%
\author{Emmanouil Kioupakis}
\affiliation{Department of Materials Science and Engineering, University of Michigan, Ann Arbor, Michigan, 48109, USA}%
\author{Feliciano Giustino}
 \affiliation{Oden Institute for Computational Engineering and Sciences, The University of Texas at Austin, Austin, Texas 78712, USA}
\affiliation{Department of Physics, The University of Texas at Austin, Austin, Texas 78712, USA}
 \email{fgiustino@oden.utexas.edu}

\date{\today}

\begin{abstract}
The past decade has seen the emergence of \textit{ab initio} computational methods for calculating phonon-limited carrier mobilities in semiconductors with predictive accuracy. More realistic calculations ought to take into account additional scattering mechanisms such as, for example, impurity and grain-boundary scattering. In this work, we investigate the effect of ionized-impurity scattering on the carrier mobility. We model the impurity potential by a collection of randomly distributed Coulomb scattering centers, and we include this relaxation channel into the \textit{ab initio} Boltzmann transport equation, as implemented in the EPW code. We demonstrate this methodology by considering silicon, silicon carbide, and gallium phosphide, for which detailed experimental data are available. Our calculations agree reasonably well with experiments over a broad range of temperatures and impurity concentrations. For each compound investigated here, we compare the relative importance of electron-phonon scattering and ionized-impurity scattering, and we critically assess the reliability of Matthiessen's rule. We also show that an accurate description of dielectric screening and carrier effective masses cam improve quantitative agreement with experiments.
\end{abstract}

\maketitle

\section{\label{sec:1} Introduction}

The ability to predict the charge transport properties of semiconductors using non-empirical \textit{ab initio} methods is of paramount importance for the design of next-generation electronics, neuromorphic computing, energy-efficient lighting, and energy conversion and storage. For example, as beyond-silicon materials for next-generation field-effect transistors are being explored, such as wide-gap semiconductors like GaN~\cite{Pushpakaran:2020}, SiC~\cite{Ramkumar:2022}, and Ga$_2$O$_3$~\cite{Green:2022}, or high-mobility materials such as GaAs~\cite{Papez:2021}, \textit{ab initio} methods for calculating transport properties with predictive accuracy are acquiring an increasingly important role.

The past decade has seen numerous developments in first-principles calculations of phonon-limited charge transport coefficients such as the electrical conductivity in metals, and the drift and Hall mobilities in semiconductors~\cite{Restrepo:2009,Li:2014,Fiorentini:2016,Kim:2016,Mustafa:2016,Ponce:2018,Protik:2020}. 
More recently, several groups turned their attention to \textit{ab initio} calculations of additional scattering mechanisms~\cite{Restrepo:2009,Caruso:2016,Lu:2019,Lu:2022,Xia:2021,Sanders:2021}. 
Among the various mechanisms, impurity scattering is of particular interest since ionized donors and acceptors are ubiquitous in high-purity doped semiconductors, and intrinsic point defects are unavoidable in all other materials~\cite{Slavcheva:2002,Callebaut:2004,Romer:2017}. In this work we focus on ionized-impurity scattering, which is expected to provide the most significant contribution to the carrier relaxation rates beyond phonons, given the long-ranged nature of the Coulomb potential.

Ionized-impurity scattering in semiconductors has first been studied via the Conwell-Weisskopf model. In this model, the scattering potential of the impurity is described using a Coulomb monopole immersed in the dielectric background of the semiconductor~\cite{Conwell:1950}. The long-range nature of this potential makes it ill-behaved at long-wavelength, and the singularity at long wavelengths is removed using an \textit{ad hoc} infrared cutoff. A better handling of this singularity is achieved in the Brooks-Herring model by considering free-carrier screening~\cite{Brooks:1955}. This latter model proved very successful~\cite{Long:1959}, and is still widely used owing to its simplicity as it only requires the electronic density of states, the carrier effective mass, the high-frequency dielectric constant, and the impurity concentration.
Further improvements upon these models were subsequently introduced, e.g., carrier statistics, dispersive electronic screening, two-impurity scattering, and atomic form factors~\cite{Kosina:1997}. While this class of models enjoyed considerable success with calculations of the carrier mobility of silicon, they do not perform as well with other semiconductors~\cite{Roschke:2001,Arvanitopoulos:2017}. These and similar other empirical adjustments make it harder to quantify the role of each scattering channels, and most importantly decrease the transferability of the models and ultimately their usefulness in materials design. 

During the past decade, considerable progress has been achieved in \textit{ab initio} calculations of charge carrier mobilities~\cite{Fiorentini:2016,Kim:2016,Ponce:2020,Restrepo:2009,Li:2014,Lu:2022}. 
These approaches are based on the use of electronic band structures from density functional theory (DFT) \cite{Hohenberg:1964,Kohn:1965}, as well as phonon dispersion relations and electron-phonon matrix elements from supercell calculations or from density-functional perturbation theory (DFPT)~\cite{Baroni:1987,Giannozzi:1991,Gonze:1997,Baroni:2001}. To achieve a numerically converged sampling of the Brillouin zone, most calculations by now employ Wannier-Fourier interpolation~\cite{Giustino:2007,Mostofi:2014,Giustino:2017}. Mobilities are then obtained by solving the \textit{ab initio} Boltzmann transport equation (\aibte) \cite{Ponce:2020}. The first study of ionized-impurity scattering from first principles was reported by Restrepo and Pantelides~\cite{Restrepo:2009}, and more recent, state-of-the-art calculations have been reported by Lu and coworkers~\cite{Lu:2022}. In this latter work, the authors find good agreement between calculated mobilities and experimental data for silicon. Additional work using a semi-empirical approach combining DFT calculations and models was also reported recently~\cite{Graziosi:2020,Ganose:2021}. 

In this work, we investigate from first principles the effect of ionized-impurity scattering on the carrier mobility of semiconductors. To this aim, we take into account both carrier-phonon and carrier-impurity scattering on the same footing, within the \aibte\ formalism as implemented in the EPW code.~\cite{Ponce:2016} Given that the shape of the impurity potential depends on the details of the crystal structure and its evaluation would require thermodynamic calculations of defects and defect levels~\cite{Freysoldt:2014}, we limit ourselves to consider the monopole term of the scattering potential and a random distribution of impurities. This simplification allows us to achieve an elegant and compact formalism, and to compute carrier mobilities by using solely the concentration of ionized impurities as input. To validate our methodology, we perform calculations for three test systems: Si, 3C-SiC, and GaP. For Si there is an abundance of experimental data and previous calculations to compare with. 3C-SiC, which is also referred to as cubic SiC or $\beta$-SiC in the literature, is considered a promising candidate for next-generation power electronics~\cite{Bhatnagar:1993,Li:2021,Li:2021:2}. Several experimental data sets are available for carrier mobility in 3C-SiC, especially for $n$-type (N) doping and less so for $p$-type doping (Al). GaP is a standard optoelectronic semiconductor which is of interest in non-linear optical switching~\cite{Zipperian:1982,Hughes:1991,Luo:2018}; experimental mobility data for GaP are available both for $n$-type doping (Sn) and $p$-type doping (Zn).
For each of these compounds we calculate the temperature-dependent carrier mobility at variable impurity concentration. We investigate the relative importance of carrier-phonon and carrier-impurity scattering, and we examine the validity of the classic Matthiessen's rule~\cite{Reif-Acherman:2015}.

The manuscript is organized as follows. In Sec.~\ref{sec:theory} we briefly summarize the \aibte\ formalism, we provide a detailed derivation of the matrix elements for carrier-impurity scattering, and we discuss the key approximations involved. In this section we also discuss free-carrier screening, and we examine under which conditions the Matthiessen rule can reliably be used in transport calculations. Section~\ref{sec:methods} is devoted to the implementation details and the calculation parameters used in this work. In Sec.~\ref{sec:results} we discuss our results for Si, SiC, and GaP. In particular, in Sec.~\ref{subsec:expcomp} we present our calculated temperature- and concentration-dependent mobilities and compare our data with experiments. In Sec.~\ref{subsec:invtau} we analyze the relative importance of phonon- and impurity-mediated scattering processes in the carrier relaxation rates. In Sec.~\ref{subsec:matt} we test Matthiessen's rule by comparing full \aibte\ calculations with the results of separate calculations including only phonon-limited or impurity-limited mobilities. In Sec.~\ref{subsec:corr} we investigate how the DFT dielectric screening and carrier effective masses influence calculated mobilities, and we test simple correction schemes along the lines of Ref.~\cite{Ponce:2018}. In Sec.~\ref{sec:conclusions} we summarize our findings and offer our conclusions. Additional details on the calculation procedure are discussed in the Appendices. 

\section{\label{sec:2} Theoretical approach}\label{sec:theory}

\subsection{Carrier mobility from the \textit{ab initio} Boltzmann transport equation}

A detailed derivation of the \aibte\ formalism is given in Ref.~\cite{Ponce:2016}. Here we limit ourselves to summarize the key equations in order to keep this manuscript self-contained. Within the linearized Boltzmann transport equation, the carrier mobility tensor is obtained as:
\begin{equation}\label{eq:BTE3}
\mu_{\alpha\beta} = -\frac{2}{\Vuc n_{\rm c}}\frac{1}{\Nuc}\sum_{n\bk}
v_{n\bk}^\alpha \partial_{E_\beta} f_{n\bk},
\end{equation}
where the factor of 2 is for the spin degeneracy, Greek indices indicate Cartesian directions, $E_\beta$ indicate the Cartesian components of the electric field, and $\partial_{E_\beta} f_{n\bk}$ is the linear variation of the electronic occupation of the state with band index $n$ and wavevector $\bk$ in response to the applied field. $v_{n\bk}^\alpha$ represents the expectation value of the velocity operator along the direction $\alpha$, for the Kohn-Sham state $n\bk$. $e$, $n_{\rm c}$, $\Vuc$, and $\Nuc$ indicate the electron charge, the carrier density, the volume of the unit cell, and the number of unit cells in the Born-von K\'arm\'an (BvK) supercell, respectively. The $n$-summation extends over all Kohn-Sham states, although in practice only those states near the chemical potential contribute to the mobility. The $\bk$-summation is over a uniform Brillouin zone grid. 

The variation $\partial_{E_\beta} f_{n\bk}$ is obtained from the self-consistent solution of the equation:
\begin{eqnarray}\label{eq:BTE1}
  && -e v^{\beta}_{n\mathbf{k}} \frac{\partial f^0_{n\mathbf{k}}}{\partial \epsilon_{n\mathbf{k}}}
  = \sum_{m\bq}\, 
     \big[ \tau^{-1}_{m\mathbf{k}+\mathbf{q} \to n\mathbf{k}} \, \partial_{E_{\beta}}f_{m\mathbf{k}+\mathbf{q}} \nonumber \\
    &&   - \tau^{-1}_{n\mathbf{k} \to m\mathbf{k}+\mathbf{q}} \, \partial_{E_{\beta}}f_{n\mathbf{k}} \big],
\end{eqnarray}
where $f^0_{n\mathbf{k}}$ denotes the Fermi-Dirac occupation of the state $n\bk$ in the absence of electric field. The quantity $\tau^{-1}_{n\mathbf{k}  \to m\mathbf{k}+\mathbf{q}}$ is the partial scattering rate from the Kohn-Sham state $n\bk$ to the state $m\bk+\bq$. In many-body perturbation theory, this rate is derived from the imaginary parts of the electron self-energy, therefore different scattering mechanisms simply add up to the lowest order in perturbation theory. In this work, we write the scattering rate as the sum of the rates of carrier-phonon scattering (ph) and carrier-impurity (imp) scattering:
\begin{equation}\label{eq:totrate}
  \displaystyle
  \frac{1}{\tau_{n\mathbf{k} \to m\mathbf{k}+\mathbf{q}}} = 
  \frac{1}{\tau^{\rm ph}_{n\mathbf{k} \to m\mathbf{k}+\mathbf{q}}} + 
  \frac{1}{\tau^{\rm imp}_{n\mathbf{k} \to m\mathbf{k}+\mathbf{q}}}. 
\end{equation}
The partial carrier-phonon scattering rate is given by~\cite{Ponce:2020}:
\begin{eqnarray}\label{eq.tau.partial}
  &&\frac{1}{\tau^{\rm ph}_{n\mathbf{k}  \to m\mathbf{k}+\mathbf{q}}} = \frac{1}{\Nuc}\sum_{\nu} \frac{2\pi}{\hbar} \left| g_{mn\nu}(\mathbf{k},\mathbf{q}) \right|^2 \nonumber \\
  &&\times \big[ (n_{\mathbf{q}\nu}+1-f^0_{m\mathbf{k}+\mathbf{q}})\delta(\epsilon_{n\mathbf{k}}\!-\!\epsilon_{m\mathbf{k}+\mathbf{q}}-\hbar\omega_{\mathbf{q}\nu}) \nonumber \\
  &&+ (n_{\mathbf{q}\nu}+f^0_{m\mathbf{k}+\mathbf{q}})\delta(\epsilon_{n\mathbf{k}}\!-\!\epsilon_{m\mathbf{k}+\mathbf{q}}+\hbar\omega_{\mathbf{q}\nu}) \big],
\end{eqnarray}
where $\epsilon_{n\bk}$ denote Kohn-Sham eigenstates, and $\omega_{\bq\nu}$ stands for the frequency of a phonon with branch index $\nu$, wavevector $\bq$, and Bose-Einstein occupation $n_{\bq\nu}$. The matrix elements $g_{mn\nu}(\mathbf{k},\mathbf{q}$) indicate the probability amplitude for the scattering of an electron from state $n\bk$ to state $m\bk+\bq$ via a phonon $\bq\nu$~\cite{Giustino:2017}. The partial rate in Eq.~\eqref{eq.tau.partial} can be obtained either from Fermi's golden rule or from many-body perturbation theory~\cite{Giustino:2017}. The carrier-impurity scattering rate required in Eq.~\eqref{eq:totrate} is derived in the next section and is given by Eq.~\eqref{eq:tauave3}.

Together, Eqs.~\eqref{eq:BTE3}-\eqref{eq.tau.partial} and \eqref{eq:tauave3} define the \aibte\ framework employed in this work. This approach consistently captures back-scattering and Umklapp processes, with a computational cost that is similar to more approximate approaches based on various relaxation-time approximations. We refer the reader to Ref.~\cite{Ponce:2020} for a comprehensive review of common approximations to the Boltzmann transport equation.

\subsection{Scattering of Carriers by ionized impurities in the monopole approximation}

To obtain the carrier-impurity scattering rate $1/\tau^{\rm imp}_{n\bk \to m\bk+\bq}$ 
we proceed as follows: (i) We derive the matrix element of the scattering potential for a single impurity in a periodic BvK supercell of the crystal unit cell; (ii) We generalize the matrix element to consider a number $N_{\rm imp}$ of impurities in the BvK supercell; (iii) From this matrix element, we obtain the scattering rate corresponding to the $N_{\rm imp}$ impurities by using the first Born approximation; (iv) We average the resulting rate over a random uniform distribution of impurity positions using a method due to Kohn and Luttinger.

\subsubsection{Scattering potential and matrix element for single impurity}

We employ the monopole approximation to describe the potential of an impurity of charge $Ze$ located at the position $\br_0$ in the BvK supercell. A more refined choice would entail explicitly calculating the impurity potential in DFT and its matrix elements. This approach was pursued in Refs.~\cite{Restrepo:2009} and \cite{Lu:2022}, but it carries the disadvantage that one needs to compute defect energetics prior to mobility calculations, and then perform rotational averages to account for the randomness of the impurity orientation. Our simpler approach is useful for systematic transport calculations when detailed knowledge of the atomic-scale structure of impurities is lacking, and can be made more accurate by incorporating dipole and quadrupole terms along the lines of Refs.~\cite{Verdi:2015,Brunin:2020,Park:2020}. 

By solving the Poisson equation in the BvK supercell and considering a background anisotropic static dielectric constant tensor $\bm\varepsilon^0 = \varepsilon^0_{\alpha\beta}$, the potential of this point charge is found to be [see Eq.~(S3) of Ref.~\cite{Verdi:2015}]:
  \begin{equation}\label{eq.S3}
    \phi({\bf r};\br_0)  = \frac{4\pi}{\Omega_{\rm sc}} \frac{Ze}{4\pi\varepsilon_0} \sum_{\bf q}
     \sum_{{\bf G}\ne -{\bf q}}\frac{e^{i({\bf q+ G})\cdot ({\bf r}-\br_0)}}
     {({\bf q}+{\bf G})\!\cdot\!\bm\varepsilon^0\!\cdot({\bf q}+{\bf G})},
  \end{equation}
modulo an inessential constant that reflects the compensating background charge. In this expression, $\varepsilon_0$ is the vacuum permittivity, $\bf G$ is a reciprocal lattice vector, and the wavevector $\bq$ belongs to a uniform Brillouin-zone grid.  Here an in the following, we consider that the BvK cell consists of $\Nuc$ unit cells, so that its volume is $\Omega_{\rm sc} = \Nuc\Vuc$, and that the Brillouin zone is discretized in a uniform grid of $\Nuc$ points. The potential $\phi(\br,\br_0)$ is periodic over the BvK supercell.

The perturbation potential resulting from this impurity is $V = \mp e \phi$ for electrons and holes, respectively. For definiteness, we consider electrons in the following. The matrix elements of the perturbation $V$ between the Kohn-Sham states $\psi_{n\bk}$ and $\psi_{m\bk+\bq}$ is given by:
  \begin{equation}\label{eq.g1}
    g_{mn}^{\rm imp}(\bk,\bq;\br_0) = \langle \psi_{m\bk+\bq} | V(\br;\br_0) | \psi_{n\bk}\rangle_{\rm sc}, 
  \end{equation}
where the integral is over the supercell. The states can be written as $\psi_{n\bk} = \Nuc^{-1/2} e^{i\bk\cdot\br} u_{n\bk}$, where $u_{n\bk}$ is the Bloch-periodic part and is normalized in the unit cell. The combination of Eqs.~\eqref{eq.S3} and \eqref{eq.g1} yields:
  \begin{equation}\label{eq.gi1}
    g_{mn}^{\rm imp}(\bk,\bq;\br_0) = \frac{-e^2}{4\pi\varepsilon_0} \frac{4\pi Z}{\Omega_{\rm sc}} 
    \sum_{{\bf G}\ne -{\bf q}}\!
    \frac{e^{-i({\bf q+ G})\cdot \br_0} B_{mn,\bG}(\bk,\bq)}{({\bf q}+{\bf G})\!\cdot\!\bm\varepsilon^0\!\cdot({\bf q}+{\bf G})}, 
  \end{equation}
having defined the overlap integral:
  \begin{equation}
    B_{mn,\bG}(\bk,\bq) = \langle u_{m\bk+\bq} | e^{i{\bf G}\cdot {\bf r}}|u_{n\bk} \rangle_{\rm uc},
  \end{equation}
which is evaluated over the unit cell. 

\subsubsection{Scattering rate from multiple impurities within the first Born approximation}

We now consider $N_{\rm imp}^{\rm sc}$ impurities located at the positions $\br_1,\br_2,\cdots,\br_{N_{\rm imp}}$ in the BvK supercell. The corresponding perturbation potential is the sum of the potentials obtained in the previous section, $V = \sum_{I=1}^{N_{\rm imp}^{\rm sc}} V(\br;\br_I)$, therefore the generalization of Eq.~\eqref{eq.gi1} to the case of multiple identical impurities reads:
  \begin{eqnarray}\label{eq.gmany}
    g_{mn}^{\rm imp}(\bk,\bq;\{\br_I\}) &=& 
    \frac{-e^2}{4\pi\varepsilon_0} \frac{4\pi Z}{\Omega_{\rm sc}} 
    \sum_{{\bf G}\ne -{\bf q}}\!
    \frac{B_{mn,\bG}(\bk,\bq)}{({\bf q}+{\bf G})\!\cdot\!\bm\varepsilon^0\!\cdot({\bf q}+{\bf G})}\nonumber \\ 
     &\times& {\sum}_{I=1}^{N_{\rm imp}^{\rm sc}} e^{-i({\bf q+ G})\cdot \br_I}. 
 \end{eqnarray}
The total scattering rate out of state $n\bk$ associated with this matrix element can be written using the first Born approximation for the scattering matrix~\cite{Sakurai:2010} [Eqs. (6.1.16) and (6.1.32)]:
\begin{equation}\label{eq:imprate0}
   \frac{1}{\tau^{\text{imp}}_{n\bk}} = 
   \sum_{m\bq} \frac{2\pi}{\hbar} |g_{mn}^{\text{imp}}(\bk,\bq;\{\br_I\})|^2 
   \delta(\epsilon_{n\bk}-\epsilon_{m\bk+\bq}).
\end{equation}
We note that this expression is an intensive quantity, as expected, i.e. it does not scale with the size of the BvK supercell [see discussion after Eq. (17)]. The partial scattering rate needed in Eq.~\eqref{eq:totrate} is then defined as:
\begin{equation}\label{eq:imprate}
     \frac{1}{\tau^{\text{imp}}_{n\bk\rightarrow m\bk+\bq}} = 
      \frac{2\pi}{\hbar} |g_{mn}^{\text{imp}}(\bk,\bq;\{\br_I\})|^2 
      \delta(\epsilon_{n\bk}-\epsilon_{m\bk+\bq}).
  \end{equation}
Unlike Eq.~\eqref{eq.tau.partial}, in this expressions we do not have the Fermi-Dirac occupations. These occupations drop out in the linearized Boltzmann transport equation, as it can be verified, for example, by setting $n_{\bq\nu}=0$ and $\omega_{\bq\nu}=0$ in Eq.~\eqref{eq.tau.partial}. In Eq.~\eqref{eq:imprate} the Dirac delta function ensures energy conservation, consistent with the fact that we are considering the scattering by a fixed potential, i.e. we are neglecting the recoil of the impurity upon collision.

By combining Eqs.~\eqref{eq.gmany} and \eqref{eq:imprate} we find:
 \begin{eqnarray}\label{eq:imprate2} &&
    \frac{1}{\tau^{\text{imp}}_{n\bk\rightarrow m\bk+\bq}}(\{\br_i\}) =  
     \frac{2\pi}{\hbar} \left[\frac{e^2}{4\pi\varepsilon_0} \frac{4\pi Z}{\Omega_{\rm sc}}\right]^2
     \delta(\epsilon_{n\bk}-\epsilon_{m\bk+\bq})
    \nonumber \\ && \times \!\!\!\!\!\!\sum_{\bf G,\bf G'\ne -{\bf q}}\!\!\!
     \frac{B_{mn,\bG}(\bk,\bq)B^*_{mn,\bG'}(\bk,\bq)}{(\bQ\cdot\!\bm\varepsilon^0\!\cdot\bQ)
     (\bQ'\cdot\!\bm\varepsilon^0\!\cdot\bQ')} 
     {\sum}_{I,J=1}^{N_{\rm imp}^{\rm sc}} e^{i(\bQ'\cdot \br_J-\bQ\cdot \br_I)}, \nonumber \\
 \end{eqnarray}
 having defined $\bQ = \bq+\bG$ and $\bQ' = \bq+\bG'$ for convenience.

\subsubsection{Kohn-Luttinger ensamble averaging of the scattering rate}

In order to account for the randomness in the distribution of impurities, we perform a configuration average of the scattering rate in Eq.~\eqref{eq:imprate2} by considering a uniform probability distribution, following the Kohn-Luttinger approach~\cite{Kohn:1957}: 
  \begin{equation}\label{eq:tauave}
    \frac{1}{\tau^{\text{imp,ave}}_{n\bk\rightarrow m\bk+\bq}} =
    \int_{\rm sc} \frac{d\br_1\cdots d\br_{N_{\rm imp}^{\rm sc}}}{\Omega_{\rm sc}^{N_{\rm imp}^{\rm sc}}}
       \frac{1}{\tau^{\text{imp}}_{n\bk\rightarrow m\bk+\bq}}
    (\{\br_i\}).
  \end{equation}
The only term that depends on the impurity positions in Eq.~\eqref{eq:imprate2} is the sum over $I,J$ on the second line. Below we evaluate the ensemble average of this sum by separating the $I=J$ and $I\ne J$ terms:
\begin{eqnarray}\label{eq:sum}
   && \hspace{-10pt}\int_{\rm sc} \frac{d\br_1\cdots d\br_{N_{\rm imp}^{\rm sc}}}{\Omega_{\rm sc}^{N_{\rm imp}}}
    {\sum}_{I,J=1}^{N_{\rm imp}^{\rm sc}} e^{i(\bQ'\cdot \br_J-\bQ\cdot \br_I)}  \nonumber \\
  && =      
   \frac{N_{\rm imp}^{\rm sc}}{\Omega_{\rm sc}}\int_{\rm sc} d\br\,
     e^{i(\bQ'-\bQ)\cdot \br} \nonumber \\
     &&
    + \frac{N_{\rm imp}^{\rm sc}(N_{\rm imp}^{\rm sc}-1)}{\Omega_{\rm sc}^2} \left[   
      \int_{\rm sc} d\br\, e^{i\bQ'\cdot \br}\right]\!\!\left[
      \int_{\rm sc} d\br\, e^{-i\bQ\cdot \br}\right]\!\!.
\end{eqnarray}
Both terms on the r.h.s.\ require the evaluation of an integral of the type:
  \begin{equation}\label{eq:int}
    \int_{\rm sc} d\br\, e^{i\bQ\cdot \br}.
  \end{equation}
This integral equals $\Omega_{\rm sc}$ for $\bQ=0$; for finite $\bQ$, we note that the integral becomes the Fourier representation of the Dirac delta when $\Nuc \rightarrow\infty$, therefore it vanishes. In this limit, Eq.~\eqref{eq:sum} reduces to:
\begin{eqnarray}\label{eq:sum2}
  &&  \int_{\rm sc} \frac{d\br_1\cdots d\br_{N_{\rm imp}}}{\Omega_{\rm sc}^{N_{\rm imp}^{\rm sc}}}
   {\sum}_{I,J=1}^{N_{\rm imp}^{\rm sc}} e^{i(\bQ'\cdot \br_J-\bQ\cdot \br_I)}  \nonumber \\
 && =      
  N_{\rm imp}^{\rm sc} \,\delta_{\bG,\bG'}
   +  N_{\rm imp}^{\rm sc}(N_{\rm imp}^{\rm sc}-1)\, 
      \delta_{\bG,-\bq}\delta_{\bG',-\bq}, \nonumber\\
\end{eqnarray}
Using Eqs.~\eqref{eq:sum2} and \eqref{eq:imprate2} inside Eq.~\eqref{eq:tauave}, we obtain:
\begin{eqnarray}\label{eq:tauave3}
 && \frac{1}{\tau^{\text{imp,ave}}_{n\bk\rightarrow m\bk+\bq}} = 
 \frac{1}{\Nuc} N_{\rm imp}^{\rm uc}
  \frac{2\pi}{\hbar} \left[\frac{e^2}{4\pi\varepsilon_0} \frac{4\pi Z}{\Vuc}\right]^2
 \nonumber \\ && \times \!\!\!\sum_{\bf G \ne -{\bf q}}
  \frac{|B_{mn,\bG}(\bk,\bq)|^2}{|(\bq+\bG)\cdot\!\bm\varepsilon^0\!\cdot(\bq+\bG)|^2}
  \delta(\epsilon_{n\bk}-\epsilon_{m\bk+\bq}),
\end{eqnarray}
where we use $N_{\rm imp}^{\rm uc} = N_{\rm imp}^{\rm sc}/\Nuc$ to denote the number of impurities per unit cell; $N_{\rm imp}^{\rm uc}$ is a dimensionless quantity.
We note that, in practical calculations, the prefactor $1/\Nuc$ in Eq.~\eqref{eq:tauave3}, which also appears in the partial carrier-phonon scattering rate in Eq.~\eqref{eq.tau.partial}, is included as a $\bk$-point weight in Brillouin zone summations, so that the sum in Eq.~\eqref{eq:BTE1} becomes $\Nuc^{-1}\sum_\bq$ and is independent of the size of the BvK supercell. 

The scattering rate given in Eq.~\eqref{eq:tauave3} is similar but not identical to alternative forms used in previous work. For example, it differs from classic approaches such as the Conwell-Weisskopf formula~\cite{Conwell:1950} and the Brooks-Herring formula~\cite{Brooks:1955} in that here the details of band structures, Kohn-Sham orbital overlaps, and anisotropic dielectric screening are fully taken into account. Furthermore, it differs from more recent \textit{ab initio} approaches such as Ref.~\cite{Restrepo:2009} in that the long-range nature of the Coulomb interaction is taken into account from the start, as opposed to being included as an \textit{ad hoc} correction. Our expression is similar to the formula provided in Ref.~\cite{Lu:2022}, except that here we take into account the periodicity of the impurity potential over the BvK supercell and the anisotropy of the dielectric tensor. The fact that we reached a similar expression as in Ref.~\cite{Lu:2022} starting from a rather different viewpoint involving the Kohn-Luttinger ensemble average lends support to both approaches.

\subsection{Free-carrier screening of the impurity potential}

The carrier-impurity scattering rate given by Eq.~\eqref{eq:tauave3} contains a singular $q^{-4}$ term that is not integrable (with $q=|\bq|$), and leads to incorrect results when used in the \aibte\ of Eq.~\eqref{eq:BTE1}. This problem was already identified by Conwell and Weisskopf \cite{Conwell:1950}, who introduced an infrared cutoff to suppress the Coulomb singularity.

The formal way to overcome this difficulty is to observe that ionized impurities are accompanied by free-carriers, which introduce metallic-like screening of the impurity potentials. In the Thomas-Fermi model, free-carriers introduce an additional screening
\begin{equation}
  \varepsilon_{\rm TF}(q) = 1+\frac{q_{\rm TF}^2}{q^2},
\end{equation}
where $q_{\rm TF}$ is the Thomas-Fermi wavevector. When used in combination with the impurity potential appearing in Eq.~\eqref{eq:tauave3}, this additional screening lifts the Coulomb singularity. In fact, by temporarily ignoring the $\bG$ vectors and the anisotropy of the dielectric tensor, free-carrier screening modifies the denominator of Eq.~\eqref{eq:tauave3} as follows:
  \begin{equation}
    \frac{1}{(\varepsilon^0 q^2)^2} \quad\xrightarrow{\hspace{10pt}}\quad
    \frac{1}{[\varepsilon_{\rm TF}(q)\varepsilon^0 q^2]^2} =
    \frac{1}{[\varepsilon^0 (q^2+q_{\rm TF}^2)]^2}, 
  \end{equation}
which tends to the finite value $1/(\varepsilon^0 q_{\rm TF}^2)^2$ at long wavelength.

To incorporate free-carrier screening in our calculations, while taking into account all details of band structures and effective masses, we employ the Lindhard dielectric function instead of the Thomas-Fermi model, following Ref.~\cite{Ashcroft:1976}. The same approach was employed in Ref.~\cite{Lu:2022}. The Lindhard dielectric function is given by:
  \begin{equation}
    \varepsilon_{\rm L}(q) = 1-\frac{e^2}{4\pi\varepsilon_0}\frac{4\pi}{q^2} \frac{2}{\Nuc \Vuc} 
         \sum_{n\bk}\frac{f^0_{n\bk+\bq}-f^0_{n\bk}}{\epsilon_{n\bk+\bq}-\epsilon_{n\bk}}.
  \end{equation}
Since the density of free-carriers is typically low in doped semiconductors, we only need the long wavelength limit of this expression. In this limit, $(f^0_{n\bk+\bq}-f^0_{n\bk})/(\epsilon_{n\bk+\bq}-\epsilon_{n\bk}) = \partial f^0_{n\bk}/\partial \epsilon_{n\bk}$, therefore we can write:
\begin{equation}
  \varepsilon_{\rm L}(q) = 1+\frac{q_{\rm TF}^2}{q^2},
\end{equation}
having introduced the effective Thomas-Fermi vector:
\begin{equation}\label{eq:qtf}
  q_{\rm TF} = \frac{e^2}{4\pi\varepsilon_0}\frac{2\cdot 4\pi}{\Nuc \Vuc} 
       \sum_{n\bk} \left|\frac{\partial f^0_{n\bk}}{\partial\epsilon_{n\bk}}\right|.
\end{equation}
For parabolic bands, Eq. (21) reduces to the Thomas-Fermi or Debye model in the respective temperature limits.
The free-carrier screening provides an additional screening mechanism to the dielectric screening of the insulating semiconductors, and is included in our calculations by replacing $\bm\varepsilon^0$ in Eq.~\eqref{eq:tauave3} by the total dielectric function:
  \begin{equation}
      \bm\varepsilon^0 \quad\xrightarrow{\hspace{10pt}}\quad \bm\varepsilon^0 + {\bf 1}\frac{q_{\rm TF}^2}{q^2},
  \end{equation}
where ${\bf 1}$ denotes the $3\times 3$ identity matrix. We note that this improved description of the screening includes temperature effects via the Fermi-Dirac occupations entering the definition of the effective Thomas-Fermi wavevector, Eq.~\eqref{eq:qtf}. 

\subsection{Matthiessen's Rule}

Matthiessen's rule~\cite{Reif-Acherman:2015} is widely employed to interpret transport measurements. In the context of carrier transport in semiconductors, this rule can be stated as follows: the contributions of different scattering channels to the mobility can be obtained by adding the reciprocals of the individual mobilities. In the case of carrier-phonon and impurity-phonon scattering, we would have:
  \begin{equation}\label{eq:matt}
    \frac{1}{\mu} = \frac{1}{\mu_{\rm ph}} + \frac{1}{\mu_{\rm imp}}.
  \end{equation}
In Sec.~\ref{sec:4} we proceed to quantify the reliability of this approximation by comparing mobility data calculated using the complete \aibte\ including both phonons and impurities with the prediction of Eq.~\eqref{eq:matt} obtained by calculating the mobility with these two scattering channels taken individually. We will show that this rule does not carry predictive power for the examples considered in this work.

From a formal standpoint, the rule expressed by Eq.~\eqref{eq:matt} is obviously related to the choice of expressing the total scattering rates as the sum or the individual rates, see Eq.~\eqref{eq:totrate}. That choice was motivated by the observation that, to first order in perturbation theory, different scattering channels do not mix. However, it is easy to see that, even when Eq.~\eqref{eq:totrate} is a good approximation, the additivity of the rates does not imply the Matthiessen rule as expressed by Eq.~\eqref{eq:matt}. To appreciate this point, we observe that the \aibte\ in Eq.~\eqref{eq:BTE1} can be recast as a linear system of the type:
 \begin{equation}
  A \times \{\partial_{E_\beta} f_{n\bk}\} = b,
 \end{equation}
 where the matrix $A$ contains the partial scattering rates $\tau^{-1}_{n\bk\rightarrow n'\bk'}$, the vector $b$ contains the drift term on the left hand side of Eq.~\eqref{eq:BTE1}, and $\{\partial_{E_\beta} f_{n\bk}\}$ denotes the vector of solutions. If we break down the matrix $A$ into its contributions from carrier-phonon and carrier-impurity scattering, $A_{\rm ph}$ and $A_{\rm imp}$ respectively, we see immediately that
 \begin{equation}
 \{\partial_{E_\beta} f_{n\bk}\} = (A_{\rm ph}+A_{\rm imp})^{-1} b \ne A_{\rm ph}^{-1}b+A_{\rm imp}^{-1} b,
 \end{equation}
therefore the additivity of the scattering rates does not imply the Matthiessen rule. This point can be made even more explicit by considering the self-energy relaxation time approximation to the \aibte. The approximation consists of neglecting the first term on the r.h.s.\ of Eq.~\eqref{eq:BTE1}, and yields the following expression for the mobility:
  \begin{eqnarray}\label{eq.serta}
    \mu_{\alpha\beta} &=& -\frac{e}{\Vuc n_{\rm c}}\frac{2}{\Nuc}
    \sum_{n\bk} \frac{\partial f_{n\bk}^0}{\partial \epsilon_{n\bk}} v_{n\bk}^\alpha v_{n\bk}^\beta
    \nonumber \\ &\times& \frac{1}{\displaystyle\frac{1}{\tau_{n\bk}^{\rm ph}}+ \frac{1}{\tau_{n\bk}^{\rm imp}}}.
  \end{eqnarray}
For this expression to be amenable to Matthiessen's rule, the scattering rates would need to be independent of the electronic state, say $\tau_{n\bk}^{\rm ph} = \tau^{\rm ph}$ and $\tau_{n\bk}^{\rm imp} = \tau^{\rm imp}$. This is typically not the case in most semiconductors. Another special case where Matthiessen's formula is meaningful occurs when one scattering mechanism dominates over the others. For example, in Eq.~\eqref{eq.serta}, when $\tau_{n\bk}^{\rm ph} \gg \tau_{n\bk}^{\rm imp}$, the expression reduces to the phonon-limited mobility. In this sense, Matthiessen's rule constitutes a simple interpolation formula between the limiting cases of phonon-limited and impurity-limited mobilities. We will analyze these aspects quantitatively in Sec.~\ref{sec:4}.

\section{\label{sec:3} Computational Methods}\label{sec:methods}

All calculations are performed using the Quantum ESPRESSO materials simulation suite~\cite{Giannozzi:2017}, the EPW code~\cite{Ponce:2016}, and the Wannier90 code~\cite{Pizzi:2020}. We employ the PBE exchange and correlation functional~\cite{Perdew:1996} and optimized norm-conserving Vanderbilt (ONCV) pseudopotentials from the PseudoDojo repository~\cite{Hamann:2013,Vansetten:2018}. For consistency with previous work, we use the experimental lattice constant of Si, SiC, and GaP at room temperature, and the plane-wave kinetic energy cutoff and quadrupole tensors reported in Ref.~\cite{Ponce:2021}. We include spin-orbit coupling for the valence bands only, to capture the splitting of the valence band top. Key calculation parameters are summarized in Tab.~\ref{tab:dft_setup}.

\begin{table}[t!]
  \centering
  \caption{Calculation parameters used in this work: Experimental lattice constant, plane wave kinetic energy cutoff, and non-vanishing elements of the quadrupole tensor are chosen to be consistent with Ref.~\cite{Ponce:2021}.}
  \begin{tabular}{l r r r}
       &  Si & 3C-SiC & GaP \\
       \hline 
Lattice constant (\AA) & 5.43 & 4.36 & 5.45 \\
Plane wave kinetic energy cutoff (eV) & 544 & 1088 & 1088 \\
$Q_{\kappa_1}$       & 11.83 & 7.41 & 13.72 \\ 
$Q_{\kappa_2}$       & -11.83 & -2.63 & -6.92 \\ 
Coarse $\mathbf k$ and $\mathbf q$ grids & 12$^3$ & 12$^3$ & 12$^3$ \\
Fine $\mathbf k$ and $\mathbf q$ electron grid & 100$^3$ & 180$^3$ & 100$^3$ \\
Fine $\mathbf k$ and $\mathbf q$ hole grid & 100$^3$ & 100$^3$ & 100$^3$ \\
       \hline
  \end{tabular}
  \label{tab:dft_setup}
\end{table}

We calculate effective mass tensors by finite differences, using a wavevector increment of $0.01\times 2\pi/a$, where $a$ is the lattice constant reported in Tab.~\ref{tab:dft_setup}. The dynamical matrix, the variations of the self-consistent potential, and the vibrational eigenfrequencies and eigenmodes are calculated using a square convergence threshold of $10^{-16}$~Ry$^2$. This threshold refers to the change of the potential variation between two successive iterations, averaged over the unit cell. Electron energies, phonon frequencies, and electron-phonon matrix elements are initially computed on a coarse wavevector mesh using the EPW code. The electron Hamiltonian, the dynamical matrix, and the electron-phonon matrix elements are then interpolated onto fine Brillouin zone grids using Wannier-Fourier interpolation~\cite{Giustino:2007,Mostofi:2014}. Long-range dipole and quadrupole corrections are employed for improved interpolation of the electron-phonon matrix elements~\cite{Verdi:2015,Sjakste:2015,Park:2020,Brunin:2020,Ponce:2021}. 

To compute carrier mobilities, only states within a narrow energy window of the band extrema are necessary. We find that, for the range of temperatures considered in this work (up to 500~K), a window of 400~meV is sufficient to obtain converged electron mobilities, and a window of 300~meV is sufficient for hole mobilities. At 300~K, converged results can be obtained by using a 200~meV window for both electrons and holes.

To evaluate the overlap matrices $B_{mn,\bG}(\bk,\bq)$ required in Eq.~\eqref{eq:tauave3} in the fine Brillouin zone grid, we follow the procedure of Ref.~\cite{Verdi:2015} and approximate them as:  
\begin{equation}
  B_{mn,\bG}(\bk,\bq) \approx \left[U(\bk+\bq) U^\dagger(\bk)\right]_{mn},
\end{equation}
where the unitary matrix $U_{mn}(\bk)$ is the diagonalizer of the interpolated Hamiltonian into the wavevector $\bk$ of the fine grid. This approximation is motivated by the fact that the carrier-impurity matrix element in Eq.~\eqref{eq:tauave3} is strongly peaked at $\bq+\bG=0$.

The Dirac delta functions appearing in Eqs.~\eqref{eq.tau.partial} and \eqref{eq:tauave3} are computed using Gaussian functions with a small broadening parameter. The results are sensitive to the choice of this parameter, therefore we accelerate the convergence by employing adaptive smearing. The procedure for the adaptive smearing of the carrier-phonon scattering rate, which involves a so-called type-III integral, is discussed in Refs.~\cite{Yates:2007,Li:2014,Ponce:2021}. The calculation of the carrier-impurity scattering rates involves instead a type-II integral of the form:
\begin{equation}
  I_{n\bk}^{\text{II}} = \sum_{m} \int \frac{d\bq}{\Omega_{\rm BZ}} f_{mn}(\bk,\bq) \,\delta(\epsilon_{m\bk+\bq}-\epsilon_{n\bk}),  
\end{equation}
where $\Omega_{\rm BZ}$ is the volume of the Brillouin zone.
In this case, adaptive broadening can be achieved by using a state-dependent width $\sigma_{m\bk+\bq}$. We follow the procedure by Ref.~\cite{Yates:2007}, which gives:
\begin{equation}\label{eq:adsmr}
  \sigma_{m\bk+\bq} = \frac{\alpha}{3}\sum_{i=1}^{3} {\bf v}_{m\bk+\bq} \cdot \frac{{\bf b}_i}{N_i},
\end{equation}
where ${\bf v}_{m \bk+\bq}$ is the band velocity, ${\bf b}_i$ is a primitive vector of the reciprocal lattice, and $N_i$ denotes the number of $\bk$-points along the direction of ${\bf b}_i$.
The coefficient $\alpha$ is a tunable parameter. 
Previous work has used $\alpha=0.29$ for electron-phonon scattering rates~\cite{Li:2014,Ponce:2021}.
We have performed a detailed converged test by comparing fixed-smearing and variable-smearing calculations, and found that values $\alpha =$~0.1-0.3 provide similar results. For simplicity, in this work we use $\alpha=0.29$ as in previous work. 

In principle we could perform calculations of carrier mobilities by setting the impurity concentration and the carrier concentration separately. This would be required, for example, for the investigation of compensation doping of semiconductors. To keep our results are general as possible, in this work we choose to focus on the simpler scenario where each impurity creates one free carrier, therefore we set the carrier density to be equal to the impurity concentration. We do not consider carrier freeze-out at low temperature, since this would require the knowledge of defect energy levels. In our calculations, the role of the carrier concentration is mainly to modulate the effective Thomas-Fermi screening wavevector in Eq.~\eqref{eq:qtf}.

\section{\label{sec:4} Results and Discussion}\label{sec:results}

\subsection{Electronic structure}\label{sec.elecst}

\begin{table}
  \centering
  \caption{
  Calculated band effective masses, band gaps, high-frequency and static dielectric constants of Si, 3C-SiC, and GaP. All calculations performed within DFT/PBE. Experimental data are from (a) \cite{Dresselhaus:1955} and \cite{Dexter:1954}, (b) \cite{Kono:1993}, (c) \cite{Bradley:1973}, (d) \cite{Kaplan:1985}, (e) \cite{Dean:1966}, (f) \cite{Collings:1980}, (g) \cite{Bimberg:1981}, (h) \cite{Lorenz:1968}, (i) \cite{Patrick:1970}, (j) \cite{Madelung:2022}, (k) \cite{Vurgaftman:2001}, (l) \cite{Kimoto:2014}. All masses are give in units of the electron mass. The band gaps are in eV. The lines tagged ``Dresselhaus'' refer to the effective masses obtained from the Dresselhaus model fitted to experimental cyclotron data, from Ref.~\cite{Dresselhaus:1955}. \vspace{5pt}}
  \begin{threeparttable}
  \begin{tabular}{lllll}
       \toprule\\[-8pt]
       This work && Si & SiC & GaP \\
      \hline\\[-8pt]
                      & $\Gamma$-X & 0.260 & 0.592 & 0.374 \\
      $m_{\text{hh}}^*$ & $\Gamma$-K & 0.550 & 1.412 & 0.837 \\
                      & $\Gamma$-L & 0.655 & 1.646 & 1.091 \\[3pt]
                      & $\Gamma$-X & 0.189 & 0.423 & 0.143 \\
      $m_{\text{lh}}^*$ & $\Gamma$-K & 0.143 & 0.328 & 0.125 \\
                      & $\Gamma$-L & 0.134 & 0.309 & 0.117 \\[3pt]
                      & $\Gamma$-X & 0.225 & 0.490 & 0.213  \\
      $m_{\text{so}}^*$ & $\Gamma$-K & 0.223 & 0.472 & 0.217 \\
                      & $\Gamma$-L & 0.214 & 0.436 & 0.206 \\[3pt]
      $m_{\text{e},\|}^{*}$ && 0.959 & 0.672 & 1.069 \\
      $m_{\text{e},\perp}^{*}$ && 0.196 & 0.230 & 0.232 \\[3pt]
      $E_{\text{g}}$  && 0.554 & 1.359 & 1.566 \\
      $\varepsilon^\infty$ && 13.00 & \phantom{0}6.93 & 10.53 \\
      $\varepsilon^0$ && 13.00 & 10.23 & 12.57 \\[2pt]
      \hline\\[-8pt]
      Experiment && Si & SiC & GaP \\
      \hline\\[-8pt]
                   & $\mathbf{B}$ along [001] & 0.46$^a$ & & \\
      $m^{*}_{hh}$ & $\mathbf{B}$ along [110] & 0.53$^a$ & & \\
                    &$\mathbf{B}$ along [111] & 0.56$^a$ & & 0.54$^c$ \\[3pt]
                    & Dresselhaus $\Gamma$-X & 0.40 & & \\
                   & Dresselhaus $\Gamma$-K & 0.56 & & \\
                     & Dresselhaus $\Gamma$-L & 0.62 & & \\[3pt]
                     & $\mathbf{B}$ along [001] & 0.171$^a$ & 0.45$^b$ & \\
      $m^*_{lh}$ & $\mathbf{B}$ along [110] & 0.163$^a$ & &  \\
                      &  $\mathbf{B}$ along [111] & 0.160$^a$ & & 0.16$^c$ \\[3pt]
                      & Dresselhaus $\Gamma$-X & 0.18 & & \\
                 &   Dresselhaus $\Gamma$-K & 0.16 & & \\
                       & Dresselhaus $\Gamma$-L & 0.15 & & \\[3pt]
      $m_{{\rm e},\|}^{*}$ && 0.97$^a$ & 0.68$^d$ & 1.15$^c$, 2.0$^k$ \\
      $m_{{\rm e},\perp}^{*}$ && 0.19$^a$ & 0.25$^d$ & 0.21$^c$, 0.25$^k$ \\[3pt]
      $E_{\text{g}}$  && 1.13$^f$ & 2.42$^g$ & 2.26$^h$ \\
      $\varepsilon^\infty$ && 11.7$^i$ & 6.52$^j$ & 9.11$^j$ \\
      $\varepsilon^0$ && 11.7$^i$ & 9.72$^j$ & 11.1$^j$ \\[2pt]
      \toprule
  \end{tabular}
  \end{threeparttable}
  \label{tab:elprop}
\end{table}

Given the importance of effective masses in mobility calculations, in this section we review briefly the band structures and effective masses of Si, SiC, and GaP. Table~\ref{tab:elprop} shows our calculated directional effective masses. Hole masses are given for the heavy-hole (hh) band, light hole (lh) band, and the spin-orbit split-off (so) band. The longitudinal ($\parallel$) and transverse ($\perp$) electron masses correspond to the principal axes of the ellipsoidal conduction band extrema.

In Tab.~\ref{tab:elprop} we see that the light hole and split-off hole masses are fairly isotropic for all compounds considered in this work. For the heavy hole masses, the $\Gamma$-X direction ([100] crystallographic direction) exhibits the lightest masses, whereas considerably heavier masses are found along the $\Gamma$-K ([110]) and $\Gamma$-L ([111]) directions. Similarly, in all compounds considered here the longitudinal electron masses are considerably heavier than the corresponding transverse masses, as expected. SiC exhibits the heaviest hole masses among SiC, GaP, and Si; while GaP exhibits the heaviest electron masses. 

Our calculated effective masses are in good agreement with previous calculations at the DFT level~\cite{Ponce:2018} as well as previous calculations at the GW level~\cite{Ponce:2018}. When comparing to experimental data, we see from Tab.~\ref{tab:elprop} that our electron effective masses are within 10\% 
of the corresponding experimental values, which is remarkable considering that we are using DFT/PBE.

In the case of the hole masses, our calculations are also in good agreement with experiments. Here we emphasize that the experimental values usually quoted are not the effective masses, but the cyclotron masses, which depend on the direction of the magnetic field and are reported in Tab.~\ref{tab:elprop}. These cyclotron masses correspond to averages of the directional masses and cannot be compared directly to DFT calculations. To extract the correct directional effective masses, in the case of silicon we used the Dresselhaus $\bk\cdot{\bf p}$ model which was fitted to experimental cyclotron data. In this model the heavy hole and light hole masses are parameterized as:
\begin{align}
\epsilon_{\text{hh}}(\bk) =& Ak^2 + [B^2k^4 + C^2(k_x^2k_y^2 + k_y^2k_z^2 + k_z^2k_x^2)]^{1/2},\\
\epsilon_{\text{lh}}(\bk) =& Ak^2 - [B^2k^4 + C^2(k_x^2k_y^2 + k_y^2k_z^2 + k_z^2k_x^2)]^{1/2},
\end{align}
where $k = |\bk|$ and the coefficients $A$, $B$, and $C$ are $-4.1\,\hbar^2/2m_{\rm e}$, $-1.6\,\hbar^2/2m_{\rm e}$,
 and $3.3\,\hbar^2/2m_{\rm e}$, respectively~\cite{Dresselhaus:1955}. From this parameterization we obtained the effective masses reported in Tab.~\ref{tab:elprop} under the keyword ``Dresselhaus''. From this table we can see that, in the case of silicon, the light hole and heavy hole masses are close to our calculated results, with the exception of the $\Gamma-X$ heavy-hole effective mass which is 65\% of the experimental value\cite{Ponce:2018}.

Our calculated dielectric constants overestimate the experimental values by 15\% at most, as expected from the underestimation of the band gaps~\cite{Patrick:1970,Madelung:2022}. 
In Sec.~\ref{subsec:corr} we discuss how one can improve the calculated mobilities by introducing \textit{a posteriori} corrections to the theoretical effective masses and dielectric constants.

\subsection{Carrier mobilities}\label{subsec:expcomp}

\subsubsection{Silicon}

Figure~\ref{fig:Si_expcomp} shows a comparison between our calculated mobilities of silicon and available experimental data, as a function of temperature and impurity concentration.
The mobilities without carrier-impurity scattering [black lines in panels (a) and (b)] decrease rapidly with temperature, as expected. We find temperature slopes (the $\beta$ in $\mu \sim T^\beta$) of $-2.1$ for electrons and $-2.4$ for holes, in agreement with previous work~\cite{Ponce:2018,Ponce:2021}. 
As we include carrier-impurity scattering, the room-temperature electron mobility of silicon reduces from 1381~\cms~to 1153~\cms\ at 1.75$\times$10$^{16}$~\pcc\ [blue line in panel (a)] and to 812~\cms~at 1.3$\times$10$^{17}$~\pcc\ [red line in panel (a)]. Similarly, the room-temperature hole mobility of silicon decreases from 600~\cms\ in the absence of impurities to 517~\cms\ for an impurity concentration of 2.4$\times$10$^{16}$~\pcc [blue line in panel (b)], and to 359~\cms\ at the impurity concentration of 2.0$\times$10$^{17}$~\cms~[red line in panel (b)].

Our calculations for the temperature-dependent electron and hole mobilities show that a single power law becomes inadequate in the presence of impurity scattering. This is also seen in the experimental data from Refs.~\cite{Morin:1954,Canali:1975,Jacobini:1977,Konstantinos:1993}, which are shown as open circles in Fig.~\ref{fig:Si_expcomp}. We note that our calculations are in good agreement with the experiments over a broad temperature range. The agreement worsens slightly at low temperature, where carrier-impurity scattering dominates. This effect likely relates to the fact that in our calculations all donors and acceptors are assumed to be fully ionized at all temperatures; as a result of this approximation, we are neglecting carrier freeze-out and hence we are likely overestimating the impurity concentration at low temperature. In Appendix~\ref{app.partialion} we show that, by taking into account the the effects of partial impurity ionization, the agreement with experiments improves at low temperature and high impurity concentration. 

Panel (c) of Fig.~\ref{fig:Si_expcomp} shows the room temperature electron mobility of silicon, as a function of impurity concentration. The electron mobility is relatively insensitive to the impurity concentration up to 10$^{16}$~\pcc. A steep decrease in the electron mobility is seen as we approach a doping density of 10$^{17}$~\pcc. Up to this concentration, our calculations (blue line) are in excellent agreement with experimental data (open black circles). Above 10$^{18}$~\pcc, while the agreement with experiment is still good, we tend to slightly overestimate the measured electron mobility. This is likely due to two effects: (i) our formalism does not take into account multiple scattering events that become important at high impurity concentration, and (ii) our calculations do not include scattering by free-carrier plasmons, which dominate the mobility at high carrier density, as shown in Refs.~\cite{Caruso:2016,Kosina:1997}. A similar overestimation was observed in Ref.~\cite{Lu:2022}.

Panel (d) of Fig.~\ref{fig:Si_expcomp} shows the room temperature hole mobility of silicon as a function of impurity concentration. As for the electrons, we find generally good agreement between calculations (blue line) and experiments (open black circles) throughout the doping range. We emphasize that the vertical scales in panels (c) and (d) are different, and that the theory/experiment deviation at high impurity concentration is similar in both panels in absolute terms. At low impurity concentration, our calculations slightly overestimate the experimental data. This effect can be ascribed to the fact that our light hole effective masses are smaller than in experiments. 

\subsubsection{Silicon carbide}

\begin{figure*}[!t]
    \centering
    \includegraphics[width=0.90\textwidth]{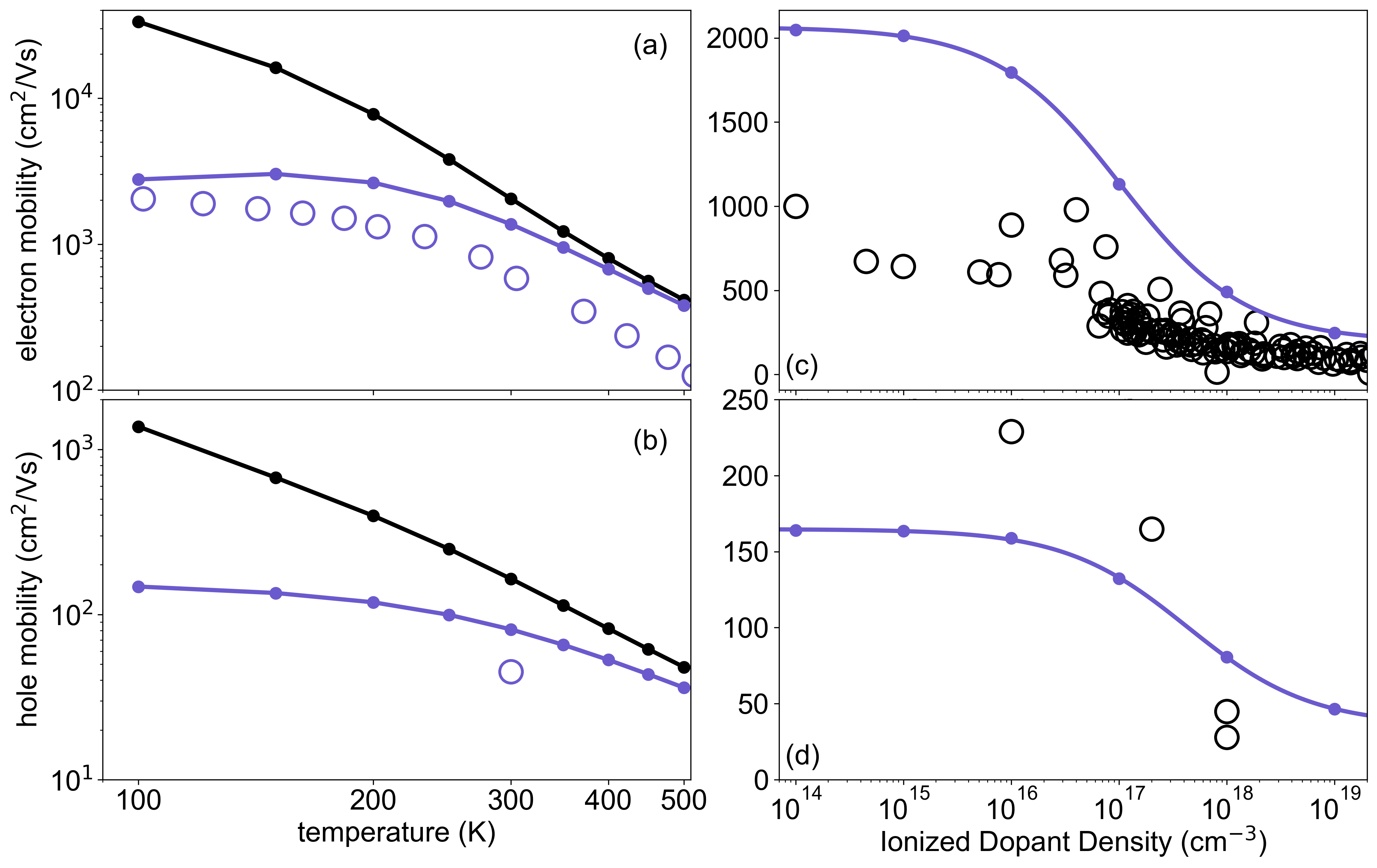}
    \caption{Comparison between our calculated carrier mobilities in 3C-SiC with experimental data. 
    (a) Electron mobility of Si as a function of temperature. The black line and symbols are phonon-limited mobilities (no impurities in the calculations); the blue line and symbols are for an impurity concentration of 5$\times$10$^{16}$~\pcc. Filled disks are calculated values, open circles are experimental data from Ref.~\cite{Shinohara:1988}.
    (b) Hole mobility of 3C-SiC as a function of temperature. The black line and symbols are phonon-limited mobilities (no impurities); the blue line and symbols are for an impurity concentration of 10$^{18}$~\pcc. All data are calculated values.
    (c) Room temperature electron mobility of 3C-SiC as a function of impurity concentration. Blue line and filled disks are calculated data, open symbols are experimental data from Ref.~\cite{Roschke:2001}, Ref.~\cite{Hirano:1995}, Ref.~\cite{Kern:1997}, and Ref.~\cite{Nelson:1966}.
    (d) Room temperature hole mobility of 3C-SiC as a function of impurity concentration. Blue line and filled disks are calculated data, open symbols are experimental data from Ref.~\cite{Wan:2002}, Ref.~\cite{Lee:2003}, Ref.~\cite{Schoner:2006}, and Ref.~\cite{Nagasawa:2008}.
    \label{fig:3CSiC_expcomp}}
\end{figure*}

In Fig.~\ref{fig:3CSiC_expcomp} we show our calculated mobilities of 3C-SiC as a function of temperature and impurity concentration, and we compare to experimental data from Refs.~\cite{Shinohara:1988,Roschke:2001,Hirano:1995,Nelson:1966,Wan:2002,Lee:2003,Schoner:2006,Nagasawa:2008}. In the case of 3C-SiC, the comparison with experiments is complicated by the high concentration of line defects that nucleate at lattice-mismatched growth substrates such as Si or 6H-SiC~\cite{Ivanov:1992,Schoner:2006}, which makes it difficult to obtain data for defect-free samples. Furthermore, most experimental data are for co-doped samples, for which the impurity and carrier concentrations are more difficult to estimate. 

In the absence of impurity scattering [black line in panel (a)], the low electron effective mass of SiC leads to very high theoretical mobilities, up to 33000~\cms\ at 100~K and up to 2000~\cms\ at room temperature. These high mobilities are in agreement with previous theoretical results~\cite{Ponce:2021}. In this case, we calculate an electron temperature exponent $\beta=-2.9$.

In panel (a) of Fig.~\ref{fig:3CSiC_expcomp} we compare our calculations (blue line) with the data reported in Ref.~\cite{Shinohara:1988} (red open circles). In that work, they synthesized 3C-SiC with $n$-type impurity density of 5.0$\times$10$^{16}$~\pcc, and obtained electron mobilities at 100~K and 300~K of 2040~\cms\ and 584~\cms, respectively. In our calculations, when we consider the same impurity concentration, we find 2773~\cms\ and 1369~\cms\ at 100~K and 300~K, respectively; therefore we overestimate the experimental data by a factor of 30\%-230\%. 

In panel (b) of Fig.~\ref{fig:3CSiC_expcomp} we show our calculated hole mobility of 3C-SiC as a function of temperature. In the absence of impurities (black line), the mobility decreases with a temperature exponent $\beta = -2.1$. In this case we could not find experimental data for uncompensated samples to compare with.
Upon including impurity scattering with an impurity concentration of 10$^{18}$~\pcc, we find a significant reduction of the mobility at low temperature (blue line), from 1373~\cms\ to 148~\cms.
At 300 K, the mobility is reduced from 165~\cms\ without impurities to 81~\cms, in good agreement with the measured value of 50~\cms\ reported in Ref.~\cite{Nagasawa:2008}. 

Panels (c) and (d) of  Fig.~\ref{fig:3CSiC_expcomp} show the room temperature electron and hole mobilities as a function of impurity concentration, respectively.
The electron mobility calculated (blue line) at low ionized donor concentration (10$^{14}$~\pcc) is 2048~\cms, and significantly overestimates the measured value 1000~\cms\ by Ref.~\cite{Hirano:1995} (open black symbols). However, our calculations get closer to experimental data in the range of concentrations above 10$^{18}$~\pcc~\cite{Kern:1997,Roschke:2001,Nelson:1966}.

The hole mobility of 3C-SiC is significantly lower than the electron mobility, as expected from much heavier hole masses shown in Tab~\ref{tab:elprop}. Our calculations (blue line) at low doping yield a mobility of 164~\cms, to be compared to 220~\cms\ measured in $p$-channel 3C-SiC devices~\cite{Lee:2003} (open black symbols). We note that the vertical scales in panels (c) and (d) differ, and that our calculated hole mobilities are in better agreement with experiment in relative terms. In particular, our data for the hole mobility fall right in the middle of the experimental trend shown in panel (d). 

\subsubsection{Gallium phosphide}

\begin{figure*}[!t]
    \centering
    \includegraphics[width=0.90\textwidth]{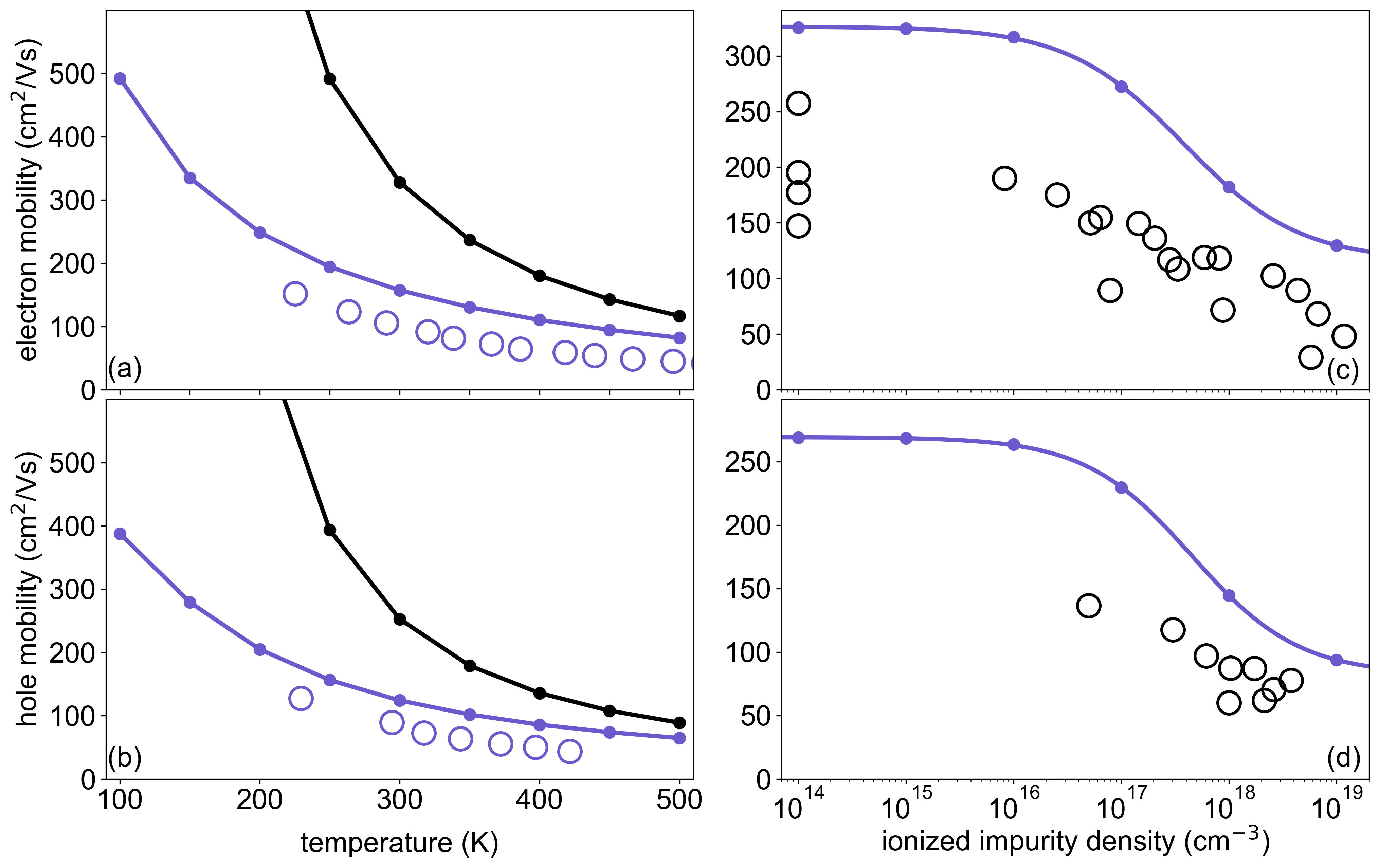}
    \caption{Comparison between our calculated carrier mobilities in GaP with experimental data.
    (a) Electron mobility of GaP as a function of temperature. The black line and symbols are phonon-limited mobilities (no impurities in the calculations); the blue line and symbols are for an impurity concentration of 2.5$\times$10$^{18}$~\pcc. Filled disks are calculated values, open circles are experimental data from Ref.~\cite{Kao:1983}.
    (b) Hole mobility of GaP as a function of temperature. The black line and symbols are phonon-limited mobilities (no impurities); the blue line and symbols are for an impurity concentration of $2\times 10^{18}$~\pcc. Filled disks are calculated values, open circles are experimental data from Ref.~\cite{Kao:1983}.
    (c) Room temperature electron mobility of GaP as a function of impurity concentration. Blue line and filled disks are calculated data, open symbols are experimental data from Ref.~\cite{Kao:1983}, Ref.~\cite{Miyauchi:1967}, Ref.~\cite{Hara:1968}, and Ref.~\cite{Craford:1971}.
    (d) Room temperature hole mobility of GaP as a function of impurity concentration. Blue line and filled disks are calculated data, open symbols are experimental data from Ref.~\cite{Kao:1983}, Ref.~\cite{Alfrey:1960}, and Ref.~\cite{Cohen:1968}.
    }
    \label{fig:GaP_expcomp}
\end{figure*}

Figure~\ref{fig:GaP_expcomp} shows our mobility calculations for GaP and a comparison with experimental data. In panel (a) we have the calculated electron mobilities as a function of temperature. In the absence of impurities, the calculated electron mobility (black line) decreases with a temperature exponent $\beta=-2.2$; the calculated mobilities at 100~K and 300~K are 4293~\cms\ and 328~\cms, respectively. 
Upon including the effect of impurity scattering (blue line), the mobility decreases significantly, reaching 157~\cms\ at room temperature for an impurity concentration of 2.5$\times$10$^{18}$~\pcc. This value is in good agreement with the measured mobility of 100~\cms\ by Ref.~\cite{Kao:1983} (blue open circles).
We note that the electron mobility of GaP is significantly lower than in silicon, despite the electron effective masses being comparable. In Sec.~\ref{subsec:invtau} we show that this effect arises from the additional polar phonon scattering that electrons experience in GaP, which is absent in silicon.  

Panel (b) of Fig.~\ref{fig:GaP_expcomp} shows the calculated phonon-limited hole mobility (black line), the mobility calculated by including impurity scattering (blue line), and experimental data (open red circles). The phonon-limited hole mobility decreases with temperature with an exponent $\beta=-2.5$.
The calculated mobilities in the absence of impurities are 5096~\cms\ and 252~\cms\ at 100~K and 300~K, respectively. Upon including impurity scattering with a concentration of 2$\times$10$^{18}$~\pcc, the mobility at room temperature decreases to 124~\cms, in good agreement with the measured value of 90~\cms\ by Ref.~\cite{Kao:1983}.

Panel (c) of Fig.~\ref{fig:GaP_expcomp} shows the room temperature electron mobility of GaP as a function of impurity concentration. In the absence of impurity scattering, we calculate a mobility of 328~\cms (blue line), which compares well with the maximum value 258~\cms\ measured in ultra-pure samples in Ref.~\cite{Miyauchi:1967} (open black symbols). In the intermediate doping regime, our calculated electron mobilities overestimate the experimental data by a factor of two~\cite{Kao:1983,Craford:1971,Hara:1968,Miyauchi:1967}, but the agreement improves at high doping levels. 

Figure \ref{fig:GaP_expcomp}(d) shows the room temperature hole mobility of GaP as a function of impurity concentration. The calculated hole mobility is 269~\cms\ at low impurity concentration, and decreases to 94~\cms\ at a concentration of 10$^{19}$~\pcc\ (blue line). Our calculations are within a factor of two from the highest measured hole mobilities across the same doping range~\cite{Kao:1983,Alfrey:1960,Cohen:1968} (open black symbols). We note that electron and hole mobilities in GaP are very similar across a wide range of temperatures and impurity concentrations (both in experiments and in our calculations), therefore GaP is an ambipolar semiconductor with well-balanced electron and hole transport.

\subsection{Carrier scattering rates}\label{subsec:invtau}

In this section we analyze and compare the scattering rates resulting from carrier-phonon and carrier-impurity processes in Si, SiC, and GaP. The Brooks-Herring model for carrier-impurity scattering~\cite{Brooks:1955}, which is based on the parabolic band approximation, predicts a scattering rate that scales as $\epsilon^{-3/2}$, where $\epsilon$ is the electron eigenvalue referred to the band extremum. This trend is a result of two competing effects: as the energy of the initial state
increases above the band bottom, the scattering phase space increases as $\epsilon^{1/2}$,
while at the same time the square modulus of the carrier-impurity matrix element given in Eq.~\eqref{eq:tauave3} decreases as $1/q^4$, which is of the order of $\epsilon^{-2}$. This simple trend
is opposite to what is expected from non-polar optical scattering and acoustic phonon scattering, which tend to increase with energy.

\begin{figure*}[!t]
    \centering
    \includegraphics[width=0.90\textwidth]{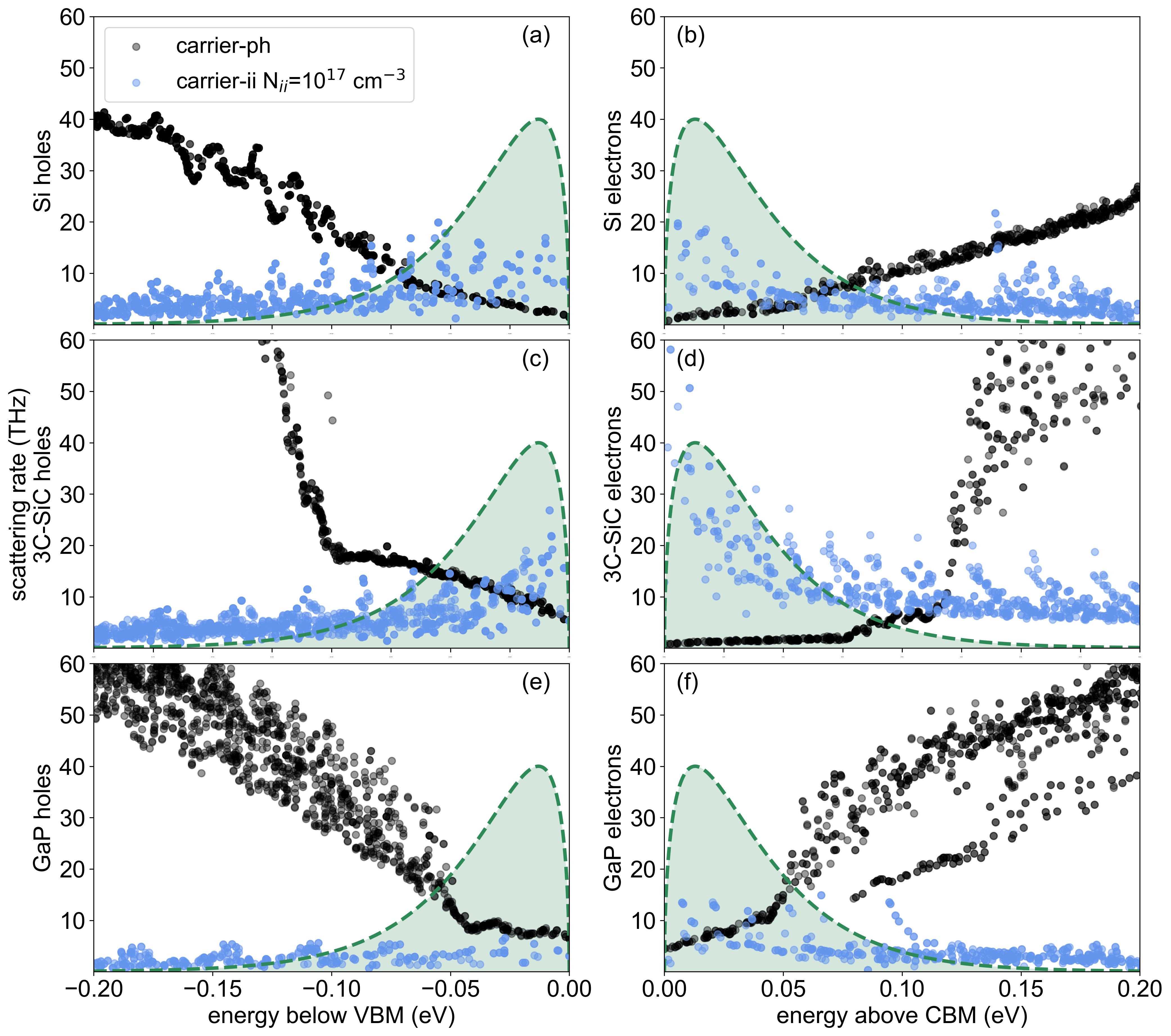}
    \caption{
    Calculated carrier scattering rates at 300~K, for an impurity concentration of $10^{17}$~\pcc.
    (a) Hole scattering rates in Si: carrier-phonon scattering rates (black disks) and carrier-impurity scattering rates (blue disks), as a function of energy referred to the valence band maximum (VBM).
    The dashed line and the shaded area represents the thermal distribution of carriers.
    (b) Electron scattering rates in Si: carrier-phonon scattering rates (black disks) and carrier-impurity scattering rates (blue disks), as a function of energy referred to the conduction band minimum (CBM).
    (c) and (d): same as (a) and (b), but for 3C-SiC.
    (e) and (f): same as (a) and (b), but for GaP.
    }
    \label{fig:invtau}
\end{figure*}

Figure~\ref{fig:invtau} shows the scattering rates $\tau^{-1}_{n\bk}$ of holes and electrons in Si [panels (a) and (b)], SiC [panels (c) and (d)], and GaP [panels (e) and (f)]. For consistency, we set the impurity concentration to 10$^{17}$~\pcc\ in all cases, which is in the middle of the range considered in Figs.~\ref{fig:Si_expcomp}-\ref{fig:GaP_expcomp}, and the temperature to 300~K. In line with the above discussion, the carrier-impurity scattering rates decrease as we move away from the band extrema, while the carrier-phonon scattering rates increase. In the two polar semiconductors that we are considering, SiC and GaP, we also see a sudden jump in the carrier-phonon scattering rates. This effect happens when the carrier energy reaches the threshold for the emission of a longitudinal optical phonon, thereby activating polar phonon scattering~\cite{Verdi:2015}.

Panels (a) and (b) of Fig.~\ref{fig:invtau} show that, in the case of silicon, the carrier-ionized-impurity scattering rates near the band edges are an order of magnitude higher than carrier-phonon rates (for an impurity concentration of 10$^{17}$~\pcc). The additional scattering by carriers causes a reduction of the mobility by $\sim 30$\% for both electrons and holes, indicating that impurity scattering is a significant effect at this impurity concentration. 
The rise of the carrier-electron scattering rates at energies around 150~meV that can be seen in panel (b) correspond to interband scattering between the two lowest conduction bands. 

Panels (c) and (d) of Fig.~\ref{fig:invtau} show the scattering rates in SiC. Unlike in silicon, here the electron and hole scattering rates differ considerably. In the case of holes, the carrier-phonon and carrier-impurity scattering rates are comparable in magnitude near the band edge, while in the case of electrons the carrier-impurity scattering dominates. This difference is reflected in the calculated mobilities, where carrier-impurity scattering reduces the phonon-limited mobility of holes by $\sim 20$\% and of electrons by $\sim 50$\% (for the impurity concentration 10$^{17}$~\pcc). 

Data for GaP are shown in panels (e) and (f) of Fig.~\ref{fig:invtau}. In this case the carrier-phonon scattering rates are comparable to the carrier-impurity scattering rates. Accordingly, the mobilities are reduced by $\sim$10\% from their values without impurity scattering. 

\subsection{Deviations from Matthiessen's Rule}\label{subsec:matt}

\begin{figure*}[!t]
    \centering
    \includegraphics[width=0.90\textwidth]{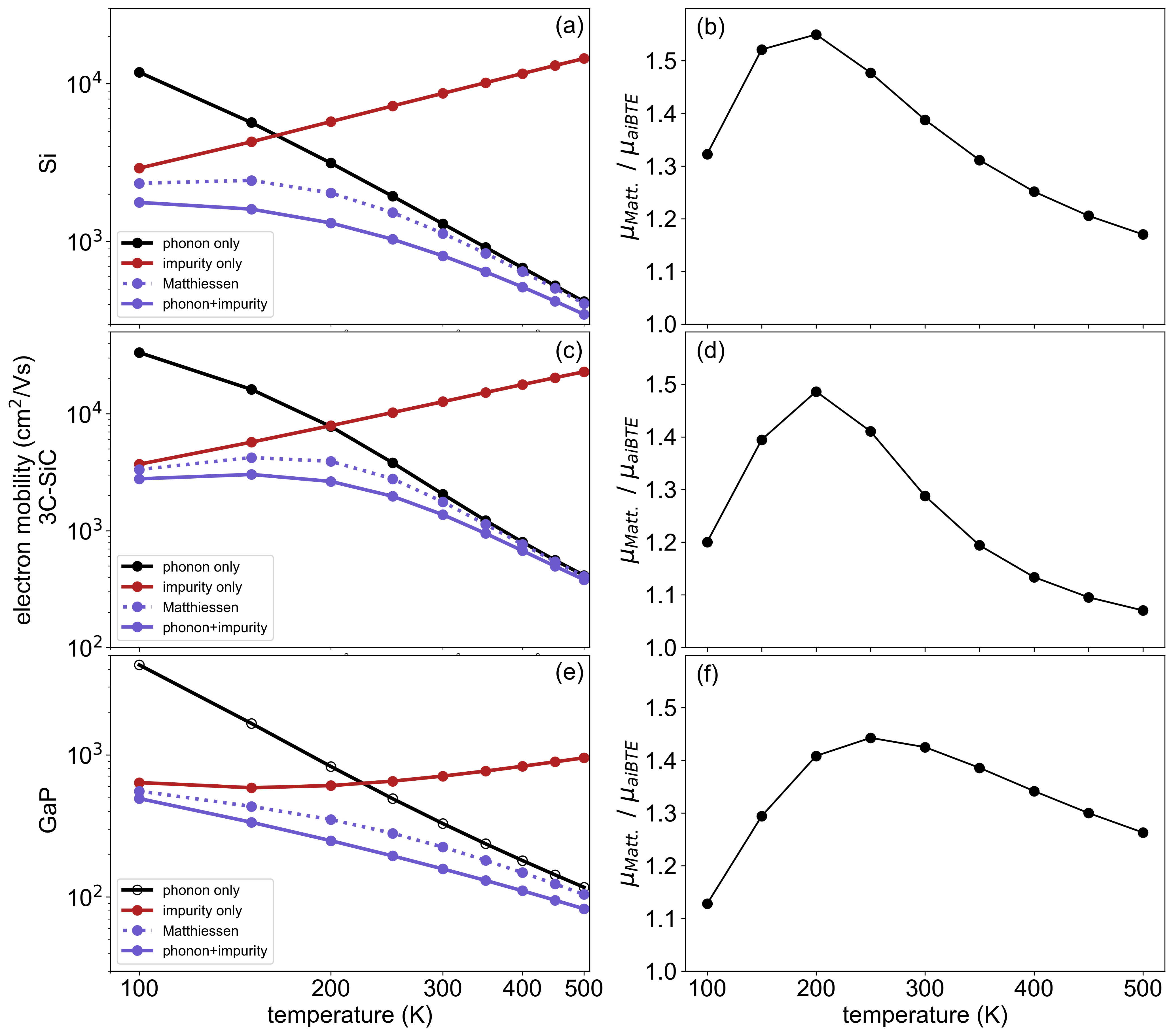}
    \caption{
    Comparison between mobility calculations performed using the \aibte\ by including both carrier-phonon and carrier-impurity scattering, and mobilities obtained by using the Matthiessen's rule.
    (a) Temperature-dependent electron mobility of Si. The black line and symbols indicate the phonon-limited mobility; the red line is the impurity-limited mobility, for an impurity concentration of $1.3\times 10^{17}$~\pcc; the dashed blue line is the mobility obtained from Matthiessen's rule; the solid blue line is the \aibte\ calculation including both phonons and impurities.
    (b) Ratio between the electron mobility of Si calculated using Matthiessen's rule and the result of the \aibte\ calculation with phonon and impurities, as a function of temperature.
    (c) and (d): Same as in (a), for for 3C-SiC with an impurity concentration of $2.5\times 10^{18}$~\pcc.
    (e) and (f):  Same as in (a), for for 3C-SiC with an impurity concentration of $5\times 10^{16}$~\pcc.    
    }
    \label{fig:matt}
\end{figure*}

In Sec.~\ref{subsec:matt} we discussed how Matthiessen's rule is formally justified only when the scattering rates are state-independent constants, or when one scattering mechanism dominates over all other mechanisms. To place that reasoning on a quantitative footing, in Fig.~\ref{fig:matt} we explicitly assess the predictive accuracy of the Matthiessen rule.

For this test, we compute the mobilities of Si, SiC, and GaP by considering the following four scenarios: (i) phonon-limited mobility $\mu_{\rm ph}$ (i.e., without including carrier-impurity scattering);
(ii) impurity-limited mobility $\mu_{\rm imp}$ (i.e., without including carrier-phonon scattering); (iii) the mobility according to Matthiessen's rule, as obtained by combining (i) and (ii) using $1/\mu_{\rm M} = 1/\mu_{\rm ph}+1/\mu_{\rm imp}$; (iv) the mobility $\mu$ calculated by including both carrier-phonon scattering and carrier-impurity scattering using the \aibte.

In panels (a), (c), and (e) we see this comparison for Si, SiC, and GaP, respectively, as a function of temperature. As expected, in all cases the phonon-limited mobilities (black lines) decrease with temperature while the impurity-limited mobilities (red lines) do increase. Their combination results into the characteristic smooth peak which is best seen in the cases of Si and SiC. In these panels, the dashed blue lines are from Matthiessen's rule, and the solid blue lines are the complete \aibte\ solutions. We see that the Matthiessen rule tends to overestimate the \aibte\ mobility, and the deviation is particularly pronounced when the phonon and impurity contributions to the mobility reduction are comparable. To quantify the deviation between \aibte\ calculations and the Matthiessen results, in panels (b), (d), and (f) of Fig.~\ref{fig:matt} we show the ratio between the two values, as a function of temperature.
In all cases we see that the use of Matthiessen's rule leads to an overestimation of the mobilities by up to 50\%, which is significant in the context of predictive calculations of transport properties. More importantly, for the compounds considered in this work (Si, SiC, and GaP), the use of Matthiessen's rule would worsen the agreement between calculated mobilities and experimental data.

Based on these findings, we caution against the use of Matthiessen's rule in future \textit{ab initio} calculations of carrier mobilities.   

\subsection{Improving the predictive power of the \aibte}\label{subsec:corr}

In this section we investigate simple approaches to improve the predictive accuracy of the \aibte\ by overcoming two standard limitations of DFT. 

The first limitation is that the DFT band gap problem typically leads to an overestimation of the dielectric screening. As a result, both carrier-phonon and carrier-impurity matrix elements tend to be underestimated in DFT~\cite{Giustino:2017,Li:2019}, and mobilities tend to be overestimated. In Ref.~\cite{Ponce:2021} it was shown that, for a set of ten semiconductors, this effect leads to mobilities which can overestimate experimental data by as much as a factor of two. To mitigate this effect, we investigate a simple scaling correction to the matrix elements as follow:
\begin{equation}
g_{mn\nu}^{\text{corr}}(\bk,\bq) = \frac{\varepsilon_{\text{DFT}}}{\varepsilon_{\text{exp}}}  g_{mn\nu}^{\text{DFT}}(\bk,\bq),
\end{equation}
where $\epsilon_{\rm DFT}$ is our calculated value, and $\epsilon_{\rm exp}$ is the experimental value. We use the high-frequency dielectric constant for the carrier-phonon matrix elements, as it was done in Ref.~\cite{Ponce:2018}, and the static dielectric constants for the carrier-impurity matrix elements [see Eq.~\eqref{eq.gmany}]. This approach is meaningful for the systems considered in this work, because the majority of scattering processes occur near the band extrema, and therefore involve small scattering wavevectors $\bq$, thus justifying the re-scaling of screening at long wavelength only. 

The second limitation of DFT calculations lies in the inaccurate curvature of the bands, which is also linked to the band gap problem, leading to slightly inaccurate carrier effective masses. This limitation could be overcome by performing GW calculations, but in this work we investigate a simpler mass scaling. 

According to Drude's formula, carrier mobilities are inversely proportional to the effective masses. Based on this observation, we consider the following scaling correction, which is directly applied to the calculated mobility:
\begin{equation}
\mu_{\text{corr}} = \frac{m^*_\text{DFT}}{m^*_\text{exp}} \mu_\text{DFT},
\end{equation}
where all masses are isotropic averages.

The three compounds considered in this work all have ellipsoidal conduction band extrema, therefore we can evaluate the average isotropic mass as follows:
\begin{equation}
m^{*} = 3 ( 1/m_\parallel^* +  2/m_\perp^* )^{-1} .
\end{equation}
Evaluating the average hole mass is more complicated owing to the band degeneracy at $\Gamma$ and the fact that experimental data usually are reported for a given magnetic field direction as opposed to a crystallographic direction (see Sec.~\ref{sec.elecst}). In the case of silicon, we evaluate the 
average mass using the values extracted from Dresselhaus' model (see Sec.~\ref{sec.elecst}). After this averaging procedure, the hole mass is calculated following Ref.~\cite{Ponce:2021}:
\begin{equation}
m^{*} = \frac{m_{\text{hh}}^{*,5/2}+m_{\text{lh}}^{*,5/2}}{m_\text{hh}^{*,3/2}+m_\text{lh}^{*,3/2}},
\end{equation}
where all quantities on the r.h.s.\ are spherical averages in k-space. In the case of SiC and GaP we are not aware of a parametrization similar to Dresselhaus', therefore we do not investigate mass corrections in these cases.

The carrier mobilities obtained by applying the above corrections are shown in Fig. \ref{fig:epscorr}. In all cases we use the experimental dielectric constants reported in Tab.~\ref{tab:elprop}.

Panels (a) and (b) show our results for silicon. The screening correction to the electron mobilities of Si reduces the calculated value at low impurity concentration from 1381~\cms\ to 1133~\cms. This reduction causes an underestimation of the experimental value by approximately 20\%. At higher impurity concentration, the corrected mobility agrees again well with experimental results. The corrections to the electron effective mass of Si are minor and do not affect the mobility.
In the case of holes, the screening and mass corrections improve considerably the agreement between theory and experiment (our calculated average hole mass is $0.43~m_{\rm e}$ while the experimental value is $0.48~m_{\rm e}$). In fact, we obtain a hole mobility of 463~\cms\ at low impurity concentration, which is within the measured value between 450 and 500~\cms\cite{Jacobini:1977,Konstantinos:1993}. The improvement is also noticeable at higher impurity concentration. 

Results for SiC are shown in panels (c) and (d) of Fig.~\ref{fig:epscorr}. In this case, we find that screening and mass corrections do not significantly improve the agreement with experiments at low impurity concentration. In particular, the screening correction reduces the electron mobility from 2047~\cms\ to 1815~\cms, and the mass correction further reduces this value to 1688~\cms. Despite these corrections, the calculated electron mobility remains too high by about a factor of two. It is possible that additional scattering mechanisms such as dislocations could contribute to reduce this difference.
In the case of the hole mobility, the screening correction reduces the calculated value at low impurity concentration from 164~\cms\ to 148~\cms, which is not significant when compared to the large spread of experimental values~\cite{Wan:2002,Lee:2003,Schoner:2006,Nagasawa:2008}. 

The screening correction appears to be successful in the case of GaP, as seen in panels (e) and (f) of Fig.~\ref{fig:epscorr}. The electron mobility at low impurity concentration reduces from 326~\cms\ to 243~\cms\ upon applying the screening correction. This value is in better agreement with the experimental data. Improved agreement with experiments is also found at higher impurity concentration. The correction to the electron effective mass of GaP is small, and as a result the change in mobility is not significant. The screening correction for holes brings the calculated data closer to the experiments. In particular, at low impurity concentration the hole mobility is reduced from 269~\cms\ to 226~\cms. 

The key takeaway from this analysis is that the screening correction to the scattering matrix elements improves the agreement between theory and experiment for the compounds considered in this work. 
Based on the above observations, we suggest that screening and mass corrections could be used for the purpose of uncertainty quantification in future \textit{ab initio} calculations of transport properties.

\section{Conclusions}\label{sec:conclusions}

In this work we have demonstrated non-empirical calculations of carrier mobilities in semiconductors using the \textit{ab initio} Boltzmann transport equations, including carrier scattering by phonons and by ionized impurities. To this end, we developed an \textit{ab initio} formalism to incorporate ionized-impurity scattering within the transport workflow based on Wannier-Fourier interpolation and implemented in the EPW code. 

We described ionized impurities by randomly distributed Coulomb scatters, and we obtained the carrier relaxation time by using the Kohn-Luttinger ensemble averaging procedure. We also incorporated the screening of the impurity potential by free-carriers, within a parameter-free effective Thomas-Fermi model. 

We validated our approach by performing an extensive set of calculations of the electron and hole mobilities of three common semiconductors, namely Si, 3C-SiC, and GaP. In all cases we find a reasonably good agreement with experimental data, except possibly for the electron mobility in SiC which is probably reduced by additional scattering at line defects in real samples. Our calculations follow closely the experimental data both as a function of temperature (at fixed impurity concentration) and as a function of impurity concentration (at fixed temperature).

Impurity scattering is found to dominate over phonon scattering at high impurity concentration and at low temperature. In the former case, the thermal distribution function of the carrier is peaked near the band edges, therefore small-$\bq$ elastic scattering by impurities dominates. In the latter case, the phonon population becomes negligible at low temperature, therefore impurities remain the only active scattering channel. These trends are fully consistent with the general understanding of carrier transport in semiconductors~\cite{Lundstrom:2009}.
We also found that the energy-dependent carrier scattering rates are strongly dependent on the detailed mechanisms at play in each compound, and vary significantly over the energy range of relevance for transport phenomena. This finding underlines the importance of detailed \textit{ab initio} calculations to achieve predictive accuracy in the description of transport phenomena of real materials.

In the presence of multiple scattering channels, it is common to analyze mobility data using the classic Matthiessen rule. However, by directly comparing \aibte\ calculations including both phonon and impurity scattering with estimates based on Matthiessen's rule, we found that the latter lead to inaccurate results, with deviations of up to 50\% with respect to \aibte\ calculations. This finding indicates that Matthiessen's rule should not be employed in predictive calculations of transport properties.

Lastly, we investigated simple corrections to DFT calculations of carrier mobilities, by scaling the calculated dielectric screening and the effective masses via their corresponding experimental values. We found that the screening correction generally improves agreement with experiments.

Overall, our present approach offers a powerful tool for calculating transport properties in a variety of semiconducting materials of immediate interest, as well as for screening new putative semiconductors in the context of materials discovery. 

Several improvements upon this work are possible. For one, we do not account for neutral impurity scattering. This additional channel could be added by generalizing our monopole model to account for dipoles and quadrupoles, following similar work performed in the context of electron-phonon interactions~\cite{Verdi:2015,Sjakste:2015,Brunin:2020,Park:2020}. Generalizations to the case of two-dimensional materials should also be possible, for example by following the related generalization of the Fr\"ohlich matrix element to two-dimensional systems~\cite{Sohier:2016,Sio:2022}. At high impurity concentration one should also account for carrier-plasmon scattering, for example as discussed in Ref.~\cite{Caruso:2016}. And of course, any improvement in the DFT band structures and electron-phonon matrix elements would be highly beneficial to further enhance the predictive power of these calculations~\cite{Li:2019}. We hope that this study will stimulate further work along these and other promising directions. 

\begin{acknowledgments}
This research is primarily supported by the Computational Materials Sciences Program funded by the U.S. Department of Energy, Office of Science, Basic Energy Sciences, under Award No. DE-SC0020129. This research used resources of the National Energy Research Scientific Computing Center, a DOE Office of Science User Facility supported by the Office of Science of the U.S. Department of Energy under Contract No. DE-AC02-05CH11231. The authors acknowledge the Texas Advanced Computing Center (TACC) at The University of Texas at Austin for providing HPC resources that have contributed to the research results reported within this paper: https://www.tacc.utexas.edu. 
\end{acknowledgments}

\appendix

\section{Incomplete ionization of dopant}\label{app.partialion}

In all calculations presented in this work, we have considered that the carrier density coincides with the impurity concentration. The implicit assumption underlying this choice is that all impurities are ionized at all temperatures. This is obviously a simplification, since the fraction of ionized impurities depends on the defect energy, the quasi Fermi level of the system, and the temperature. These aspects have already been discussed in the case of silicon in Ref.~\cite{Lu:2022}.

In this Appendix we analyze the effect of incomplete ionization for the case of silicon.
To estimate the fraction $f$ of ionized impurities at a given temperature, we use the Fermi-Dirac distribution evaluated at the defect level of the impurity atom, $\epsilon_{\rm d}$~\cite{Ashcroft:1976}:
\begin{equation}
f = \frac{1}{N_{\text{imp}}^{\text{uc}}} \sqrt{N_{\text{imp}}^{\text{uc}} \sum_{n,\bk} \frac{1}{e^{(\epsilon_{n\bk}-\epsilon_d)/k_B T}+1}}.
\end{equation}
Here, $N_{\text{imp}}^{\text{uc}}$ is the number of impurities per unit cell. This fraction vanishes when the temperature goes to zero, and approaches unity at high temperature.

\begin{figure*}[!t]
    \centering
    \includegraphics[width=0.90\textwidth]{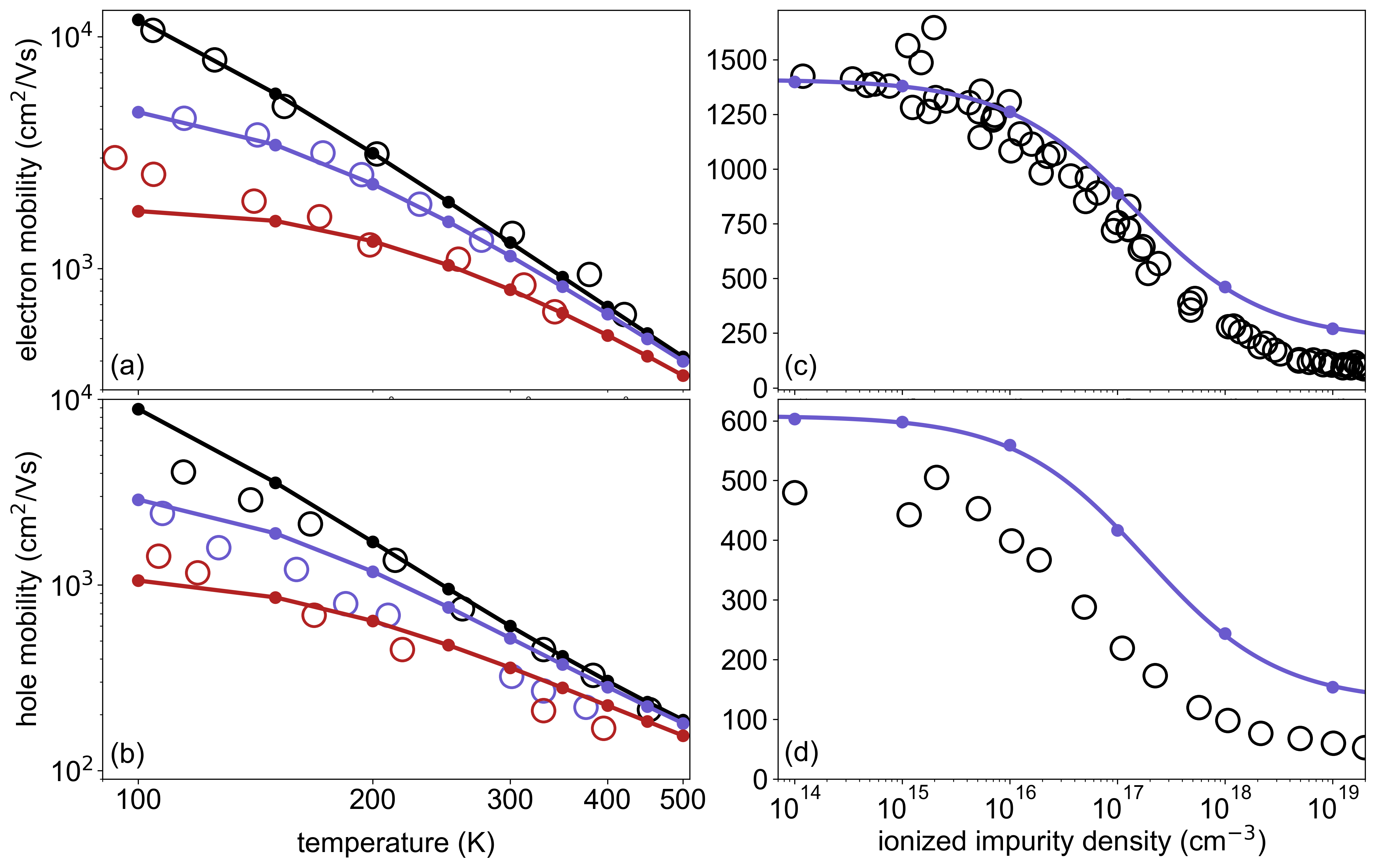}
    \caption{Comparison between our calculated carrier mobilities in Si with experimental data. 
    (a) Electron mobility of Si as a function of temperature. The black line and symbols are for low impurity concentration (no impurities in the calculations;  $<10^{12}$~\pcc\ impurities in the experiment); the blue line and symbols are for an impurity concentration of 1.75$\times$10$^{16}$~\pcc; the red line and symbols are for a concentration of 1.3$\times$10$^{17}$~\pcc. Filled disks are calculated values, open circles are experimental data from Ref.~\cite{Canali:1975} (black) and \cite{Morin:1954} (blue and red).
    (b) Hole mobility of Si as a function of temperature. The black line and symbols are for low impurity concentration (no impurities in the calculations; 10$^{12}$~\pcc\ impurities in the experiment); the blue line and symbols are for an impurity concentration of 2.4$\times$10$^{16}$~\pcc; the red line and symbols are for a concentration of 2.0$\cdot$10$^{17}$~\pcc. Filled disks are calculated values, open circles are experimental data from Ref.~\cite{Ottaviani:1975} (black) and \cite{Morin:1954} (blue and red).
    (c) Room temperature electron mobility of Si as a function of impurity concentration. Blue line and filled disks are calculated data, open black circles are experimental data from Ref.~\cite{Jacobini:1977}.
    \label{fig:Si_expcomp}}
\end{figure*}

\begin{figure*}[!t]
    \centering
    \includegraphics[width=0.90\textwidth]{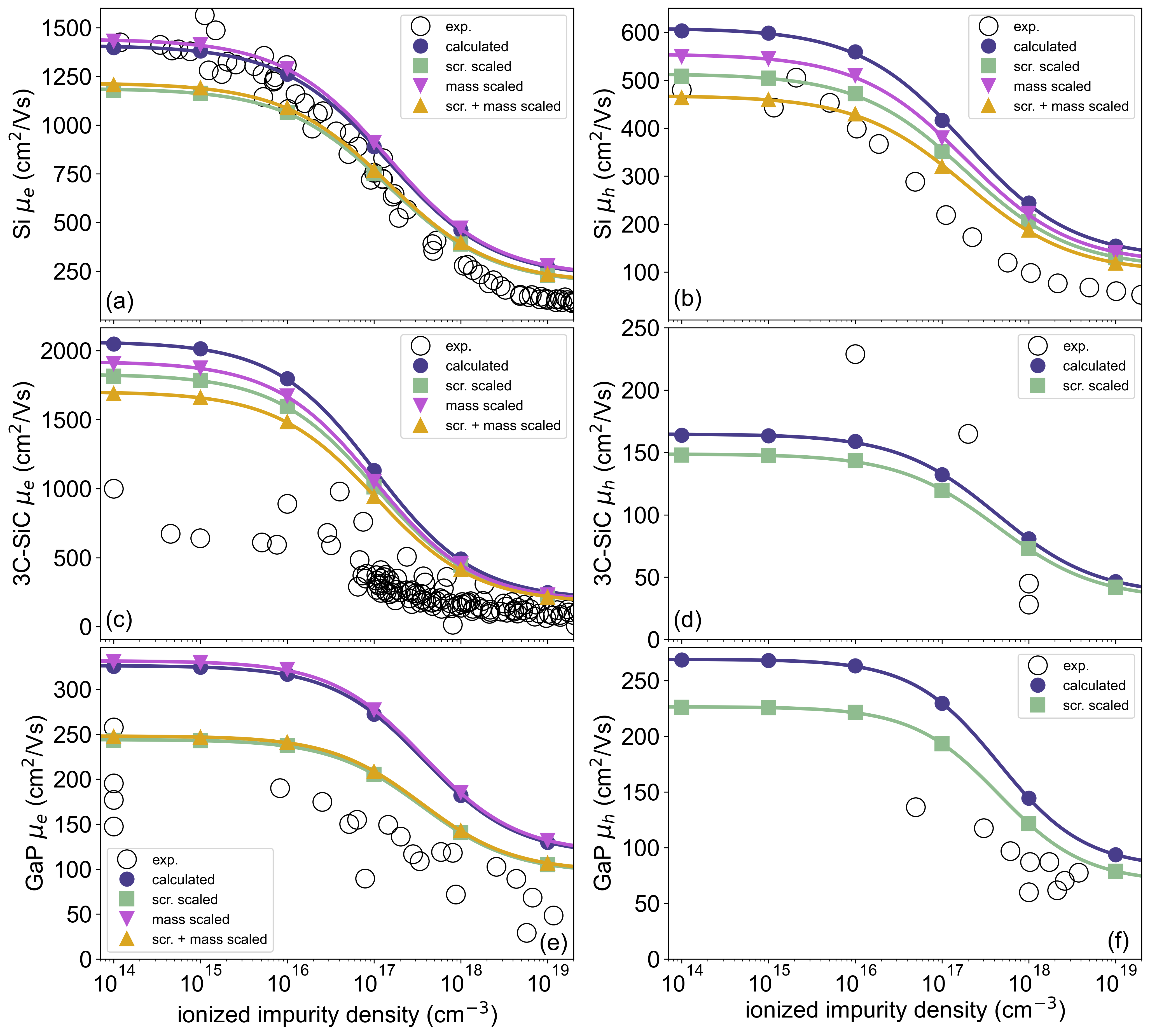}
    \caption{
    Comparison of correction schemes for improving the predictive accuracy of \aibte\ calculations of mobilities.
    (a) Room-temperature electron mobility of Si, as a function of impurity concentration. Blue lines and disks indicate the uncorrected \aibte\ results; green lines and disks indicate calculations with matrix elements corrected for screening; purple lines and disks are calculations corrected for the effective masses; yellow lines and disks include corrections for both the screening and the effective masses. Open black circles are experimental data.  
    (b) Room temperature hole mobility of Si as a function of impurity concentration: uncorrected (blue); with screening correction (green); with effective mass correction (purple); and with both screening and mass correction (yellow). 
    (c) and (d): Same as in (a) and (b) but for 3C-SiC.
    (e) and (f): Same as in (a) and (b) but for GaP.
    The experimental data are the same as those reported in Figs.~\ref{fig:Si_expcomp}, \ref{fig:3CSiC_expcomp}, and \ref{fig:GaP_expcomp}~\cite{Jacobini:1977,Roschke:2001,Hirano:1995,Kern:1997,Nelson:1966,Wan:2002,Lee:2003,Schoner:2006,Nagasawa:2008,Kao:1983,Craford:1971,Hara:1968,Miyauchi:1967,Alfrey:1960,Cohen:1968}.}
    \label{fig:epscorr}
\end{figure*}

\begin{figure*}[!t]
    \centering
    \includegraphics[width=0.49\textwidth]{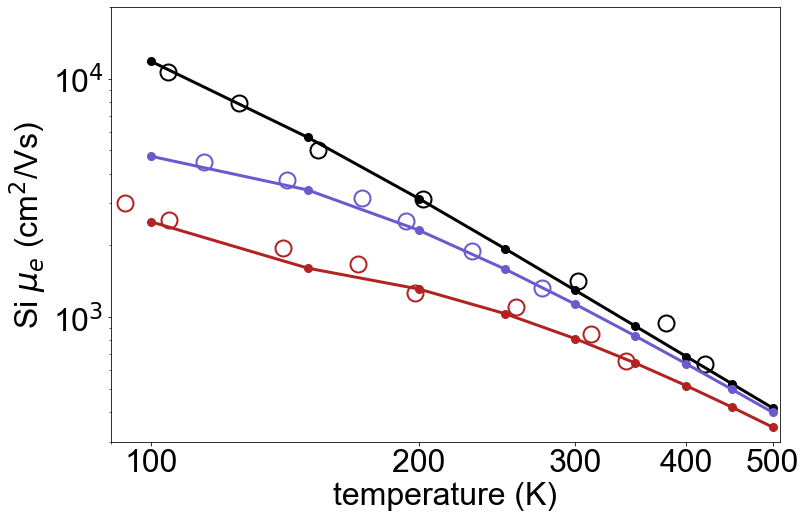}
    \caption{Electron mobility in Si as a function of temperature, including the effect of incomplete ionization of the dopants. The black line and disks are the calculated phonon-limited mobilities. These data are compared to measurements for pristine silicon (impurity concentration $<10^{12}$\pcc), from Ref.~\cite{Canali:1975}.
    The blue disks and line are calculations for an impurity concentration of $1.75\times 10^{16}$~\pcc, taking into account incomplete dopant ionization as described in Appendix~\ref{app.partialion}.
    Experimental data are from Ref.~\cite{Morin:1954}.
    The red disks and line are for an impurity concentration of $1.3\times 10^{17}$~\pcc, 
    taking into account incomplete dopant ionization.
    Experimental data are from Ref.~\cite{Morin:1954}.
    }
    \label{fig:part}
\end{figure*}

In Fig.~\ref{fig:part} we show the influence of incomplete impurity ionization on the electron mobility of Si. In these calculations, we used $\epsilon_{\rm d} = $45~meV as measured from the conduction band bottom~\cite{Wu:2017}. By comparing these curves with Fig.~\ref{fig:Si_expcomp}(a), we see that the effect of incomplete ionization improves the agreement with experiments at low temperature and high doping (red curves in both figures). This is precisely the range where carrier-impurity scattering tends to dominate over phonon scattering, therefore it is important to have a precise determination of the impurity concentration in this range. A more systematic assessment of these effects will require a broader database of experimental mobilities to compare with.

\clearpage\newpage

\bibliography{apssamp}

\end{document}